\documentclass[12pt,dvips]{article}
\usepackage{rotating}
\usepackage{amsmath}
\usepackage{mathtools}
\usepackage{amsthm}
\usepackage{amsfonts}
\usepackage[dvips]{epsfig}
\usepackage{graphicx}
\usepackage{amssymb}
\usepackage{cancel}
\usepackage{pstricks} 
\usepackage{lscape}
\usepackage{cancel}
\usepackage{slashed}
\usepackage{pstricks} 
\usepackage{lscape}
\usepackage{color}
\usepackage{cite}
\newcommand{\beq}{\begin{equation}}
\newcommand{\eeq}{\end{equation}}
\newcommand{\ga}{\lower.7ex\hbox{$\;\stackrel{\textstyle>}{\sim}\;$}}
\newcommand{\la}{\lower.7ex\hbox{$\;\stackrel{\textstyle<}{\sim}\;$}}

\newcommand{\kahler}{K\"ahler }

\begin{document}

\def\thefootnote{\fnsymbol{footnote}}

\begin{flushright}
{\tt KCL-PH-TH/2015-23}, {\tt LCTS/2015-13}, {\tt CERN-PH-TH/2015-122}  \\
{\tt ACT-04-15, MI-TH-1513} \\
{\tt UMN-TH-3438/15, FTPI-MINN-15/26} \\
\end{flushright}

\vspace{0.7cm}
\begin{center}
{\bf {\large Calculations of Inflaton Decays and Reheating:} \\
\vspace{0.1in}
with Applications to No-Scale Inflation Models}
\vspace {0.1in}
\end{center}

\vspace{0.05in}

\begin{center}{
{\bf John~Ellis}$^{a}$,
{\bf Marcos~A.~G.~Garcia}$^{b}$,
{\bf Dimitri~V.~Nanopoulos}$^{c}$ and
{\bf Keith~A.~Olive}$^{b}$
}
\end{center}

\begin{center}
{\em $^a$Theoretical Particle Physics and Cosmology Group, Department of
  Physics, King's~College~London, London WC2R 2LS, United Kingdom;\\
Theory Division, CERN, CH-1211 Geneva 23,
  Switzerland}\\[0.2cm]
  {\em $^b$William I. Fine Theoretical Physics Institute, School of Physics and Astronomy,\\
University of Minnesota, Minneapolis, MN 55455, USA}\\[0.2cm]
{\em $^c$George P. and Cynthia W. Mitchell Institute for Fundamental Physics and Astronomy,
Texas A\&M University, College Station, TX 77843, USA;\\
Astroparticle Physics Group, Houston Advanced Research Center (HARC), \\ Mitchell Campus, Woodlands, TX 77381, USA;\\
Academy of Athens, Division of Natural Sciences,
Athens 10679, Greece}\\
\end{center}

\bigskip

\centerline{\bf ABSTRACT}

\noindent  
{\small We discuss inflaton decays and reheating in no-scale Starobinsky-like models of inflation, calculating the effective
equation-of-state parameter, $w$, during the epoch of inflaton decay, the reheating
temperature, $T_{\rm reh}$, and the number of inflationary e-folds, $N_*$, comparing analytical
approximations with numerical calculations. We then illustrate these results with applications to models based on no-scale
supergravity and motivated by generic string compactifications, including scenarios where the 
inflaton is identified as an untwisted-sector matter field with direct Yukawa couplings to MSSM fields,
and where the inflaton decays via gravitational-strength interactions.
Finally, we use our results to discuss the constraints on these models imposed by present measurements of the 
scalar spectral index $n_s$ and the tensor-to-scalar perturbation ratio $r$, converting them into constraints on
$N_*$, the inflaton decay rate and other parameters of specific no-scale inflationary models.}

\vspace{0.2in}

\begin{flushleft}
May 2015
\end{flushleft}
\medskip
\noindent

\newpage

\section{Introduction}

A new generation of experiments on the cosmic microwave background (CMB),
particularly Planck \cite{planck15} and experiments searching for $B$-mode polarization, is providing
detailed probes of models of cosmological inflation. In particular, recent data from Planck
provide a very precise measurement of the scalar spectral index $n_s$. 
Recent polarization results from the Planck and BICEP2 experiments~\cite{planckbicep} have focused attention
on models~\cite{reviews} that predict relatively low values of the scalar-to-tensor perturbation ratio $r$,
and a next generation of $B$-mode polarization experiments is expected to produce
results soon.
Examples of low-$r$ models include the
Starobinsky model based on a $R + R^2$ extension of the Einstein Lagrangian~\cite{Staro,MC,Staro2},
and related models such as Higgs inflation~\cite{Higgsinf}, which typically predict $r \sim 0.003$.
This is considerably below the current upper limit $r \lesssim 0.08$~\cite{planck15,planckbicep}, and models predicting
values of $r$ that are significantly larger than in the Starobinsky model may also be
compatible with the data.

We expect that the framework for physics at the Planck scale and below should be
supersymmetric~\cite{ENOT,nost,hrr,gl1}. In addition to the myriad motivations from particle physics,
supersymmetry also renders technically natural the fact that the magnitude of the
CMB perturbations is small, by ensuring that radiative corrections to the
requisite small mass scale and/or field coupling(s) are under control. The appropriate
supersymmetric framework for cosmology is supergravity, but generic supergravity
models of inflation soon encountered problems \cite{eta}. It was therefore proposed to
consider models of inflation~\cite{gl2,KQ,EENOS} based on no-scale supergravity~\cite{no-scale,LN}~\footnote{We
recall that compactifications of string theory lead generically to no-scale supergravity models~\cite{Witten1985},
adding to their appeal.}, which are capable of mitigating
these problems~\footnote{For some alternative supergravity-based models, see~\cite{enq,bg,msy2,Davis:2008fv,Ant,ADEH,klr,klor,lln,bww}.}.

Following the 2013 Planck data release, three of us re-examined~\cite{ENO6,ENO7,ENO8}
no-scale models of inflation based on a K\"ahler potential of the form
\begin{equation}
K \; = \; -3\ln\left(T+T^* - \frac{|\phi|^2}{3} \right) \, ,
\label{noscaleK}
\end{equation}
where $T$ can be identified with the string compactification modulus
and $\phi$ is a generic matter field. With suitable choices of superpotential 
$W (T, \phi)$, no-scale models can reproduce Starobinsky-like
predictions with the inflaton identified as either the compactification modulus or a matter field,
thanks to their conformal equivalence to $R + R^2$ gravity. There has
subsequently been an outburst of interest in these and related no-scale models of 
inflation~\cite{FKR,buch,Pallis,adfs,FeKR,EGNO,EGNO2,EGNO3,EGNO4,Li,KT,bdhw,klt,twyy,LT,df,roest}.
In particular, we have shown how no-scale supergravity could accommodate models
interpolating between the Starobinsky and chaotic quadratic models of inflation,
analyzing their predictions for $n_s$ and $r$ as functions of the number of e-folds
during inflation, $N_*$, including also two-field effects~\cite{EGNO2,EGNO3}.

We have recently studied various phenomenological aspects of such no-scale
models of inflation, stressing how they could be embedded in compactifications
of string theory~\cite{EGNO4}. We analyzed possible string assignments for the inflaton and matter
fields, as well as mechanisms for supersymmetry breaking, inflaton couplings
and decays. We showed that different no-scale supergravity models led to
different estimates of the reheating temperature after inflation, $T_{\rm reh}$, and found a 
connection between the reheating temperature and the possible mechanism
of supersymmetry breaking.

The emerging data on $n_s$ and $r$ are beginning to impose interesting
constraints on the number of inflationary e-folds $N_*$, which depends on
$T_{\rm reh}$ and the equation of state during the epoch of
inflaton decay, which is conveniently characterized by the effective
equation-of-state parameter $w_{\rm int}$~\cite{twyy,MRcmb,reheatconst}. The cosmological data are therefore
starting to impose supplementary constraints on inflationary models that
may help discriminate among no-scale scenarios, casting some
light on the mechanism of inflaton decay, and possibly supersymmetry breaking.

In this paper we study these connections in some detail, comparing analytic and numerical
calculations in Section~2 and evaluating 
$w_{\rm int}$, $T_{\rm reh}$ and hence $N_*$. In Section~3 we apply these results in various 
Starobinsky-like no-scale inflationary
models, and in Section~4 we use the CMB bounds on $n_s$ and $r$ to constrain $N_*$
and thereby parameters in scenarios for no-scale inflation. Section~5 summarizes our results and discusses
future prospects. We plan in a subsequent paper to study the low-energy constraints on 
supersymmetry breaking, and their complementary implications for 
no-scale models of inflation.

\section{On the Number of e-Folds in No-Scale Inflation}

In the slow-roll approximation and assuming entropy conservation after reheating,
the number of e-folds to the end of inflation can be expressed as~\cite{LiddleLeach,MRcmb,planck15}
\beq
N_* = 66.9 - \ln\left(\frac{k_*}{a_0H_0}\right) + \frac{1}{4}\ln\left(\frac{V_*^2}{M_P^4\rho_{\rm end}}\right) + \frac{1-3w_{\rm int}}{12(1+w_{\rm int})}\ln\left(\frac{\rho_{\rm reh}}{\rho_{\rm end}}\right) - \frac{1}{12}\ln g_{\rm reh}\ ,
\label{howmany}
\eeq
where $k_*$ is the wave number  at the reference scale, $a_0$ and $H_0$ are the present 
cosmological scale factor and Hubble expansion rate, respectively, 
$V_*$ is the inflationary energy density at the reference scale, 
$\rho_{\rm end}$ and $\rho_{\rm reh}$ are the energy densities at the end of inflation and after reheating, 
$w_{\rm int}$ is the {\it e-fold} average of the equation-of-state parameter during the thermalization epoch, 
and $g_{\rm reh}$ is the number of equivalent bosonic degrees of freedom after reheating:
$\rho_{\rm reh} = (\pi^2/30)g_{\rm reh}T_{\rm reh}^4$.

We now discuss the evaluations of the quantities appearing in (\ref{howmany}),
with an initial focus on Starobinsky-like models of inflation that we extend later to related no-scale models.

\subsection{The Inflationary Energy Density $\boldsymbol{V_*}$}

The Starobinsky potential $V = \frac{3}{4}m^2 M_P^2 (1 - {\rm exp}^{- \sqrt{\frac{2}{3}} \frac{\phi}{M_P}})^2$ 
(where $M_P \equiv 1/\sqrt{8 \pi G_N} \simeq 2.4 \times 10^{18}$~GeV is the reduced Planck mass)
is nearly scale-invariant for large values of the inflaton field $\phi$:
for $\phi\gg M_P$, $V\simeq \frac{3}{4}m^2M_P^2$. This value is therefore a good first approximation to $V_*$. 
We can refine this value by recalling that the number of e-folds of inflation may be calculated in the slow-roll approximation as 
\begin{align}
N_* &\simeq -\frac{1}{M_P^2}\int_{\phi_*}^{\phi_{\rm end}}\frac{V}{V'}\,d\phi\\
&= \frac{\sqrt{6}}{4M_P}(\phi_{\rm end}-\phi_*) - \frac{3}{4}\left(e^{\sqrt{\frac{2}{3}}\frac{\phi_{\rm end}}{M_P}} - e^{\sqrt{\frac{2}{3}}\frac{\phi_*}{M_P}}\right)\, ,\label{Ns2}
\end{align}
where $\phi_*$ and $\phi_{\rm end}$ are the values of the inflaton field at the reference scale $k_*$
and the end of inflation, respectively, and the prime denotes differentiation with respect to $\phi$.
Equation (\ref{Ns2}) may be inverted to obtain $\phi_*$ in terms of the lower Lambert function $W_{-1}(x)$.
In practice, the asymptotic form $W_{-1}(x)=\ln(-x)-\ln(-\ln(-x))+\cdots$ is sufficient to obtain a good estimate for $\phi_*$, namely
\begin{align}\label{phistar}
\phi_* 
%
&\simeq \sqrt{\frac{3}{2}}M_P \ln\left[\frac{4}{3}N_*-\sqrt{\frac{2}{3}}\frac{\phi_{\rm end}}{M_P}+e^{\sqrt{\frac{2}{3}}\frac{\phi_{\rm end}}{M_P}}\right]\,.
\end{align}
This in turn implies that
\beq\label{Vstar1}
V_* \simeq \frac{3}{4}m^2M_P^2 \left(1- \frac{3}{ 4N_*-\sqrt{6}\frac{\phi_{\rm end}}{M_P}+3e^{\sqrt{\frac{2}{3}}\frac{\phi_{\rm end}}{M_P}} }\right)^2.
\eeq
In the range $50<N_*<70$, this yields $0.728m^2M_P^2< V_* <0.734 m^2M_P^2$, a
result that is in good agreement with the more exact values that we obtain from numerical integration of the equations of motion.

The mass of the scalar field is not arbitrary, but is determined from the amplitude of the scalar power spectrum. 
At horizon crossing, the amplitude may be evaluated in the slow-roll approximation to be
\beq
A_{S_*} \simeq \frac{V_*^3}{12\pi^2M_P^6(V_*')^2} = \frac{3}{8\pi^2}\left(\frac{m}{M_P}\right)^2\sinh^4\left(\frac{\phi_*}{\sqrt{6}M_P}\right)\,.
\eeq
Using the approximation (\ref{phistar}), this relation may be inverted to solve for the mass of the inflaton field,
\beq
m\simeq 8\pi M_P\sqrt{\frac{2A_{S*}}{3}}\,\frac{ \frac{4}{3}N_*-\sqrt{\frac{2}{3}}\frac{\phi_{\rm end}}{M_P}+e^{\sqrt{\frac{2}{3}}\frac{\phi_{\rm end}}{M_P}} }{ \left(\frac{4}{3}N_*-\sqrt{\frac{2}{3}}\frac{\phi_{\rm end}}{M_P}+e^{\sqrt{\frac{2}{3}}\frac{\phi_{\rm end}}{M_P}}-1\right)^2 }\,.
\eeq
In the range $50<N_*<70$ and using $\ln(10^{10}A_{S*}) = 3.094$ \cite{planck15}, this corresponds to
\beq
1.218 \; < \; 10^5(m/M_P) \; < \; 1.464 \, .
\label{inflatonmass}
\eeq
Substitution in (\ref{Vstar1}) leads to our final expression for the energy density at horizon crossing,
\beq
V_*\simeq \frac{18\pi^2A_{S*}M_P^4}{\left(N_*-\sqrt{\frac{3}{8}}\frac{\phi_{\rm end}}{M_P} + \frac{3}{4}(e^{\sqrt{\frac{2}{3}}\frac{\phi_{\rm end}}{M_P}}-1)\right)^2}\,,
\label{Vstar}
\eeq
which we use in our subsequent analysis.

\subsection{The Energy Density $\boldsymbol{\rho_{\rm end}}$}

In the case of single-field inflation, the evolution of the homogeneous, canonically-normalized scalar $\phi$ in the presence of a 
spatially-flat Friedmann-Robertson-Walker geometry is governed by the equations
\begin{align}
\ddot{\phi}+3H\dot{\phi}+V'(\phi)&=0\,,\label{frw1}\\
\frac{1}{2}\dot{\phi}^2+V(\phi)&=3M_P^2H^2\,, \label{frw2}
\end{align}
where $H$ is to the Hubble parameter. Differentiating (\ref{frw2}) with respect to time
and substituting (\ref{frw1}) yields the relation 
\beq\label{frw3}
\dot{H}=-\frac{\dot{\phi}^2}{2M_P^2}\,.
\eeq
Using (\ref{frw3}), the time dependence can be eliminated from the Friedmann equation, which leads to the Hamilton-Jacobi form of the equations of motion,
\begin{align}
[H'(\phi)]^2 - \frac{3}{2M_P^2}H(\phi)^2 & = -\frac{1}{2M_P^4}V(\phi)\,,\\
\dot{\phi} &= -2M_P^2H'(\phi)\,.
\end{align}
The {\it Hubble} slow-roll parameters are defined by
\begin{alignat}{2}
\epsilon_H(\phi) & \equiv 2M_P^2\left(\frac{H'(\phi)}{H(\phi)}\right)^2 &&= \epsilon_1\,,\\
\eta_H(\phi) & \equiv 2M_P^2\frac{H''(\phi)}{H(\phi) } &&=\epsilon_1-\frac{\epsilon_2}{2}\,,
\end{alignat}
where $\epsilon_{1,2}$ are the first and second Hubble flow-functions, $\epsilon_1 \equiv -\dot{H}/H^2$, 
$\epsilon_{i+1}\equiv \dot{\epsilon_i}/(H\epsilon_i)$~\cite{Leach:2002ar,Finelli:2009bs}.
In terms of these parameters, the condition for inflation to occur is precisely
\beq
\ddot{a}>0\ \Longleftrightarrow\ \epsilon_{H}<1\,,
\eeq
which implies that inflation ends when $\epsilon_H=1$.

Alternatively, one can consider the conventional {\it potential} slow-roll parameters
\begin{align}
\epsilon_V (\phi)& \equiv \frac{M_P^2}{2}\left(\frac{V'(\phi)}{V}\right)^2\,,\\
\eta_V(\phi) & \equiv M_P^2 \left( \frac{V''(\phi)}{V} \right)\,,
\end{align}
which are fully determined by the shape of the inflationary potential. They can be expressed in terms of the slow-roll parameters via the relations
\begin{align}
\epsilon_V &= \epsilon_H\left(\frac{3-\eta_H}{3-\epsilon_H}\right)^2\,,\label{epsV}\\
\eta_V &= (2M_P^2\epsilon_H)^{1/2}\frac{\eta_H'}{3-\epsilon_H} + \left(\frac{3-\eta_H}{3-\epsilon_H}\right)(\epsilon_H+\eta_H)\,,\label{etaV}
\end{align}
which show that $\epsilon_V=1$ is only a first-order approximation at the end of inflation.
It can be shown that the first term in (\ref{etaV}) is of higher order in slow roll \cite{Liddle:1994dx}. Neglecting this term, we can eliminate $\eta_H$ from equations (\ref{epsV}, \ref{etaV}) at the end of inflation, to obtain
\beq
\text{End of inflation:} \qquad \epsilon_V \simeq (1 + \sqrt{1 - \eta_V/2})^2 \, ,
\eeq
which can be used to calculate $\phi_{\rm end}$. 

This equation involves the scalar potential and its first two derivatives,
and can be solved in closed form in the case of a power-law potential $V=a(\phi/M_P)^n$, yielding
\beq
\text{Power-law:} \qquad \phi_{\rm end} \; \simeq \; \left(\frac{2n-1}{2\sqrt{2}}\right)M_P \,.
\eeq
This deviates from the exact result found by numerical integration of the equations of motion (\ref{frw1},\ref{frw2}) by less than 5\% for $n\geq1$. In the case of the Starobinsky potential, in a leading-order analytic approximation the end of inflation is reached when 
\beq
\text{Starobinsky:} \qquad \phi_{\rm end} \simeq \sqrt{\frac{3}{2}}\ln\left(\frac{2}{11}(4 + 3\sqrt{3})\right) M_P \; \simeq \; 0.630M_P \,,
\eeq
which is to be compared to the more exact value $\phi_{\rm end}=0.615M_P$ obtained by the numerical integration of the Friedmann and Klein-Gordon equations. 

The energy density at the end of inflation may then be obtained in a straightforward way by noting that the slow-roll parameter  $\epsilon_H$ can be rewritten as $\epsilon_H=\frac{3}{2}(1+w)$, where $w \equiv p/\rho$ is the equation-of-state parameter. When inflation ends, $w=-1/3$, which implies 
\beq
\dot{\phi}_{\rm end}^2=V(\phi_{\rm end}) \, .
\eeq
In the two cases discussed above, this may be evaluated to obtain
\beq
\rho_{end}\simeq 
\begin{cases}
\text{Power-law:} \qquad & \frac{3a}{2}\left(\frac{2n-1}{2\sqrt{2}}\right)^n\ , \\
\text{Starobinsky:} \qquad & \frac{9}{8} \left(1 - \frac{11}{2(4 + 3 \sqrt{3})}\right)^2 m^2M_P^2 \; \simeq \; 0.182 m^2M_P^2\ .
\end{cases}
\eeq
The latter can be compared with $\rho_{\rm end}=0.175m^2M_P^2$,
which is obtained if we use the exact result for the Starobinsky potential,
corresponding to the Hubble parameter $H_{\rm end}=0.242 \ m$.

\subsection{The Energy Density at Reheating $\boldsymbol{\rho_{\rm reh}}$}

We calculate the energy density at reheating assuming that the inflaton decay is perturbative, 
with a rate $\Gamma_{\phi}$.  As a first approximation, one can consider the decay to be 
complete when $\Gamma_{\phi}=t^{-1}$. However, as
we will see in Fig.~\ref{fig:womega}, in general the decay of the inflaton is incomplete at this time. Instead, we assume
here that reheating is complete when the bulk of the energy density is provided by the relativistic decay products of the inflaton:
\beq\label{omegareh}
\Omega_{\gamma} \equiv \frac{\rho_{\gamma}}{\rho_{\phi}+\rho_{\gamma}} = 1 - \delta\,,
\eeq
for some suitable $\delta \ll1$. 

\begin{figure}[h!]
\centering
	\scalebox{0.5}{\includegraphics{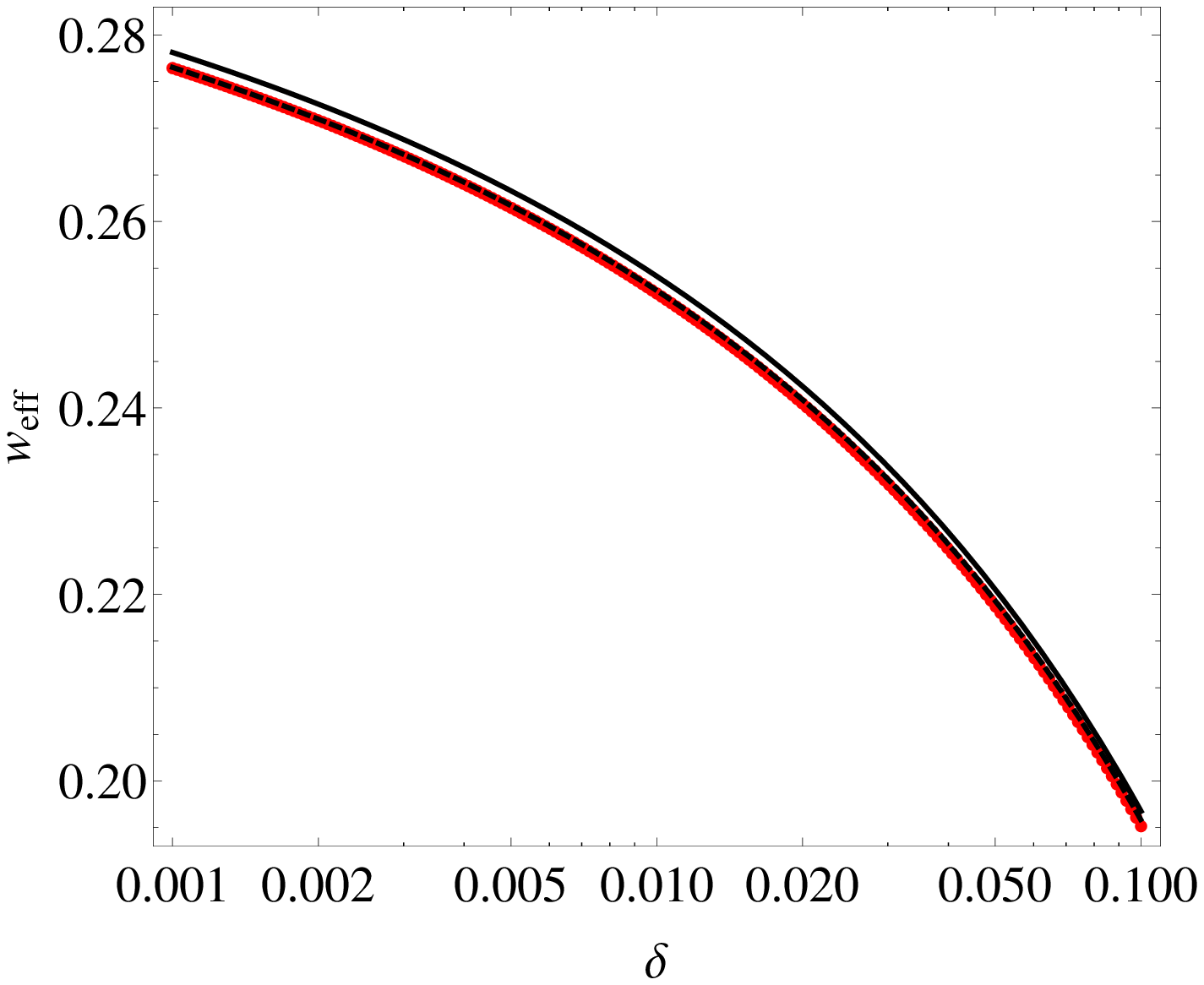}} 
	\scalebox{0.5}{\includegraphics{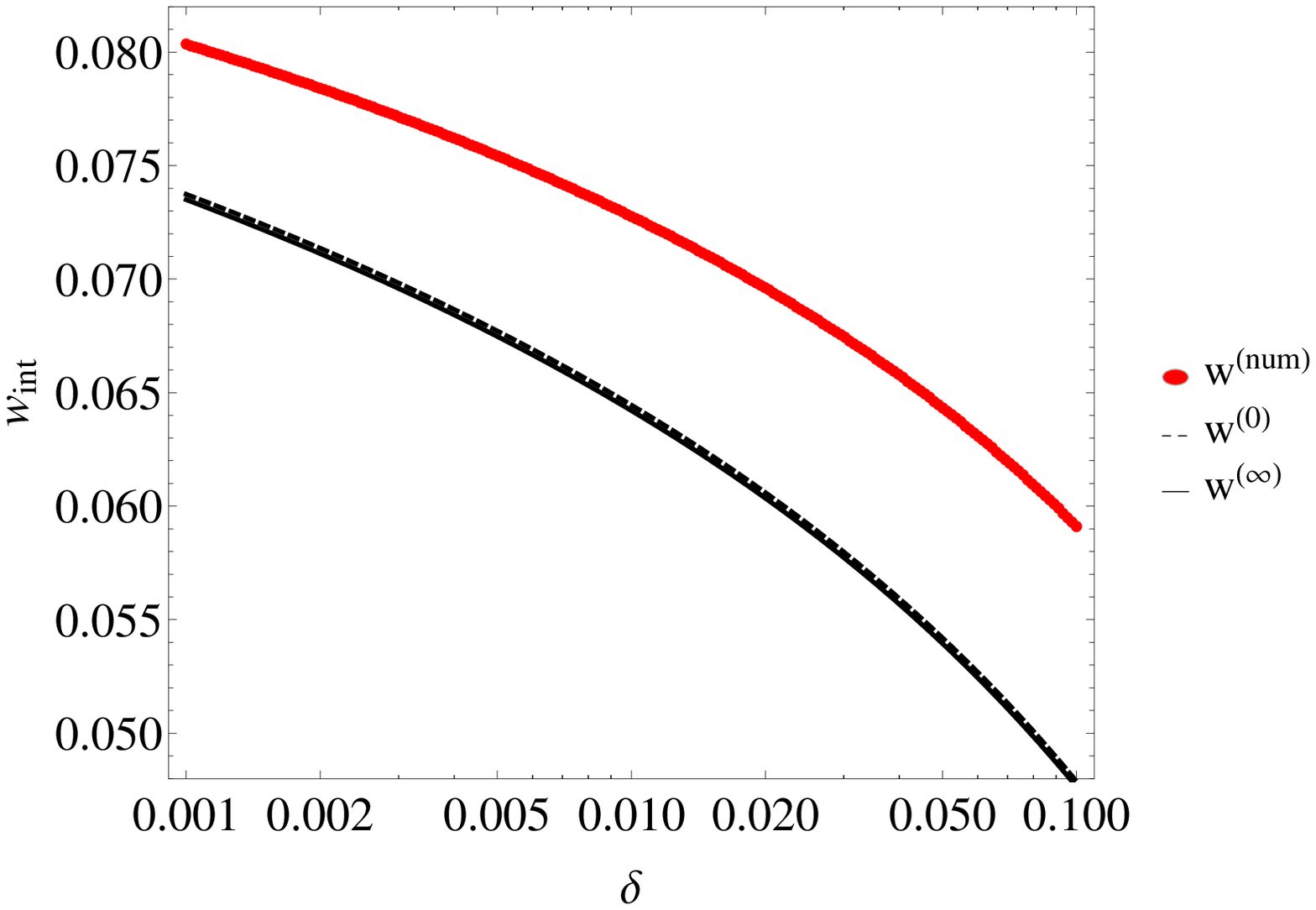}} \\
	\caption{\it The effective equation-of-state parameters for $\Gamma_{\phi} = 10^{-4} m$
	as functions of the end of reheating defined by the parameter $\delta$ in (\protect\ref{omegareh}). 
	The numerical data are shown as red points, while the first-order and iterated approximations discussed
	in the text are displayed as dashed and solid lines, respectively. 
	Left panel: The time average $w_{\rm eff}$. Right panel: The e-fold average $w_{\rm int}$.}
	\label{fig:wdelta}
\end{figure}

During reheating, the evolution of the inflaton field $\phi$ and the relativistic decay products can be described by the equations
\begin{align}
\ddot{\phi}+3H\dot{\phi} + \Gamma_{\phi}\dot{\phi}+V'&=0 \, , \label{eom1}\\
\dot{\rho}_{\gamma} + 4H\rho_{\gamma} &= \Gamma_{\phi}\rho_{\phi} \, , \label{eom2}\\
\rho_{\phi}+\rho_{\gamma} &= 3M_P^2H^2 \, . \label{eom3}
\end{align}
It is only after integration of these equations that the moment when the decay is complete can be computed.
However, we can find an approximate value when $m \gg \Gamma_\phi$ by averaging over the scalar field oscillations. 
The average energy density of the inflaton then corresponds to $\langle \rho_{\phi}\rangle = \langle \dot{\phi}^2/2\rangle + \langle V\rangle \simeq \langle \dot{\phi}^2\rangle $, and the average equation of motion (\ref{eom1}) simplifies to 
\begin{align}
\dot{\rho}_{\phi} + 3H\rho_{\phi} &= -\Gamma_{\phi}\rho_{\phi} \, . \label{aeom1}
\end{align}
These equations have the solution
\begin{align} \label{rho1a}
\rho_{\phi}(t) &=\rho_{\rm end}\left(\frac{a(t)}{a_{\rm end}}\right)^{-3}e^{-\Gamma_{\phi}(t-t_{\rm end})}\,,\\
\rho_{\gamma}(t) &=\rho_{\rm end}\left(\frac{a(t)}{a_{\rm end}}\right)^{-4}\int_{\Gamma_{\phi}t_{\rm end}}^{\Gamma_{\phi}t} \left(\frac{a(t')}{a_{\rm end}}\right)e^{u_{\rm end}-u}\, du\,, \label{rho2a}
\end{align}
where $u=\Gamma_{\phi}t'$, and we have assumed that the energy density of all relativistic 
degrees of freedom is negligible until the end of inflation \cite{Turner:1983he}. 
If the decay rate is small, the reheating epoch spans a considerable interval of time, 
and $t_{\rm reh}\gg t_{\rm end}$. In this limit, the scale factor and the Hubble parameter during the reheating epoch can be approximated as
\beq\label{aH_reh}
a(t)\simeq a_{\rm end} \left(\sqrt{\dfrac{3}{4}\rho_{\rm end}} (1+w)t/M_P\right)^{\frac{2}{3(1+w)}}\, ,\quad H\simeq \frac{2}{3(1+w)t}\,.
\eeq
If we approximate $w$ by its {\em time}-averaged value during reheating,
\beq
w_{\rm eff} \equiv \frac{1}{t_{\rm reh}-t_{\rm end}}\int_{t_{\rm end}}^{t_{\rm reh}} w(t)\,dt\,,
\eeq
then we can compute $t_{\rm reh}$ by iteration. Consider as a first approximation $w=0$, corresponding to the scalar field oscillations in the absence of decay. In this case, the solution (\ref{rho1a}),(\ref{rho2a}) can be combined with the constraint (\ref{omegareh}) to yield
\beq
\begin{aligned}
\delta^{-1}-1 &= e^{\Gamma_{\phi}t_{\rm reh}}(\Gamma_{\phi}t_{\rm reh})^{-2/3}\int_0^{\Gamma_{\phi}t_{\rm reh}}u^{2/3}e^{-u}\, du\\
 &= e^{\Gamma_{\phi}t_{\rm reh}}(\Gamma_{\phi}t_{\rm reh})^{-2/3}\boldsymbol{\gamma}({\textstyle \frac{5}{3}},\Gamma_{\phi}t_{\rm reh}) \, ,
\end{aligned}
\label{relation}
\eeq
where in this context $\boldsymbol{\gamma}$ denotes the lower incomplete gamma function. The relation (\ref{relation}) can be inverted numerically for any value of $\delta<1$. For $\delta<10^{-1}$, the solution may be approximated by
\beq\label{treh_app0}
\Gamma_{\phi}t_{\rm reh}^{(0)} \simeq 0.754-1.113\ln\delta \, ,
\eeq
where the upper index denotes the degree of the approximation. An estimate for $w_{\rm eff}$ may be derived by noting that the equation of state evaluated over the scalar field oscillations is
just one third of the fraction of the total density provided by the relativistic decay products of the inflaton:
\beq\label{womega}
\langle w\rangle  = \frac{\frac{1}{2}\langle\dot{\phi}^2\rangle-\langle V\rangle +\frac{1}{3}\langle \rho_{\gamma}\rangle}{\frac{1}{2}\langle\dot{\phi}^2\rangle+\langle V\rangle+\langle\rho_{\gamma}\rangle} \simeq \frac{\rho_{\gamma}/3}{\rho_{\phi}+\rho_{\gamma}} = \frac{1}{3}\Omega_{\gamma}\,.
\eeq
Therefore, the zeroth order approximation to the time average of $w$ can be calculated as 
\beq\label{weff}
w_{\rm eff}^{(0)} \approx \frac{1}{3\Gamma_{\phi}t_{\rm reh}}\int_0^{\Gamma_{\phi}t_{\rm reh}} \frac{ \boldsymbol{\gamma}({\textstyle \frac{5}{3}},u) }{ \boldsymbol{\gamma}({\textstyle \frac{5}{3}},u) +u^{2/3}e^{-u} }\,du \simeq 0.271\,,
\eeq
where for illustration purposes we have considered the end of reheating to occur when $\delta=0.002$ in (\ref{omegareh}).
The functional dependence of $w_{\rm eff}^{(0)}$ on $\delta$ is illustrated in the left panel of Fig. \ref{fig:wdelta} for the case of an inflaton
decay rate $\Gamma_{\phi} = 10^{-4} m$.

\begin{figure}[h!]
\centering
	\scalebox{0.8}{\includegraphics{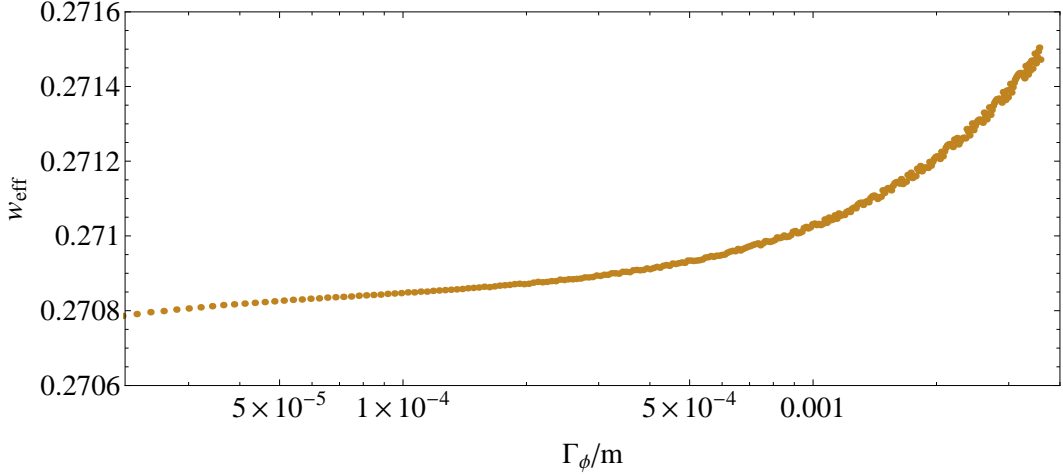}}
	\caption{\it The effective equation-of-state parameter $w_{\rm eff}$ as a function of the decay rate of the inflaton. The data converge to the value $w_{\rm eff}=0.271$ for $\Gamma_{\phi}\ll m$. }
	\label{fig:weff}
\end{figure}

The computed value of $w_{\rm eff}^{(0)}(\delta)$ may be substituted in (\ref{aH_reh}) to calculate a first-order approximation
to the reheating time,
$t_{\rm reh}^{(1)}$, which may in turn be used to evaluate $w_{\rm eff}^{(1)}$, and so on. This iterative procedure relaxes after a few steps, resulting in
\beq\label{treh_app}
\Gamma_{\phi}t_{\rm reh}^{(\infty)} \simeq 0.655-1.082\ln\delta \, ,
\eeq
and $w_{\rm eff}^{(\infty)}(0.002)\simeq 0.273$. We have checked that the iterative solution is not
sensitive to the initial choice of $w$. We see in the left panel of Fig.~\ref{fig:wdelta} that $w_{\rm eff}^{(\infty)}$,
shown as the solid black line,
agrees very well with the results of integrating numerically the evolution equations, shown as the red points.
Numerical results for $w_{\rm eff}$ are shown in Fig.~\ref{fig:weff}
as functions of the decay rate of the inflaton. 
We see that the results converge to the value $w_{\rm eff}\approx 0.271$ for $\Gamma_{\phi}\ll m$, 
a result consistent with the approximation (\ref{weff}).

The energy density at the end of reheating may then be approximated by
\beq\label{rhoreh}
\rho_{\rm reh}=3M_P^2H_{\rm reh}^2 \simeq 3M_P^2\left(\frac{2}{3(1+w_{\rm eff})t_{\rm reh}}\right)^2 =  \frac{4}{3}(1+w_{\rm eff})^{-2}M_P^2\Gamma_{\phi}^2 (0.655-1.082\ln\delta)^{-2}\,.
\eeq
The corresponding reheating temperature $T_{\rm reh}$ (assuming rapid thermalization \cite{therm}) is given by
\beq
T_{\rm reh} \; = \; \left( \frac{30 \rho_{\rm reh}}{\pi^2 g_{\rm reh}} \right)^{1/4} \, ,
\label{Treh}
\eeq
where the number of degrees of freedom $g_{\rm reh}$ would be 915/4 for $T_{\rm reh}$ above all the sparticle masses ${\tilde m}$ and
falling at lower $T_{\rm reh}$, e.g., to $g_{\rm reh} = 427/4$ for $m_t < T_{\rm reh} < {\tilde m}$.

\subsection{Result for $\boldsymbol{N_*}$}

Using the previous results, we can rewrite (\ref{howmany}) for Starobinsky-like models in the form
\begin{align}
N_* &= 68.66 - \ln\left(\frac{k_*}{a_0H_0}\right) + \frac{1}{4}\ln\left(A_{S*}\right) - \frac{1}{2} \ln \left( N_*-\sqrt{\frac{3}{8}}\frac{\phi_{\rm end}}{M_P} + \frac{3}{4}e^{\sqrt{\frac{2}{3}}\frac{\phi_{\rm end}}{M_P}} \right) \notag \\  
&\quad + \frac{1-3w_{\rm int}}{12(1+w_{\rm int})}\big(2.030+2\ln\left(\Gamma_{\phi}/m\right) -2\ln (1+w_{\rm eff}) - 2\ln(0.655-1.082\ln\delta) \big) \label{res1}\\
&\quad - \frac{1}{12}\ln g_{\rm th}\, . \notag
\end{align}
If we define
\begin{align}
\mathcal{N}_1 &=  68.66 - \ln\left(\frac{k_*}{a_0H_0}\right) + \frac{1}{4}\ln\left(A_{S*}\right)  - \frac{1}{12}\ln g_{\rm th} \notag\\
&\quad + \frac{1-3w_{\rm int}}{12(1+w_{\rm int})}\big(2.030+2\ln\left(\Gamma_{\phi}/m\right) -2\ln (1+w_{\rm eff}) - 2\ln(0.655-1.082\ln\delta) \big)\,, \\
\mathcal{N}_2 &= -\sqrt{\frac{3}{8}}\frac{\phi_{\rm end}}{M_P} + \frac{3}{4}e^{\sqrt{\frac{2}{3}}\frac{\phi_{\rm end}}{M_P}} \simeq 0.86\,,
\end{align}
then (\ref{res1}) can be inverted in terms of the (upper) Lambert function $W_0$, resulting in
\begin{align} \label{NstarW}
N_* & = \frac{1}{2}W_0\left(2e^{2(\mathcal{N}_1+\mathcal{N}_2)}\right) - \mathcal{N}_2\\
& = \mathcal{N}_1 + \frac{1}{2}\ln 2 - \frac{1}{2}\ln\left(2(\mathcal{N}_1+\mathcal{N}_2)+\ln2\right) + \cdots
\end{align}
which is the basis for our subsequent analysis. 

This expression for $N_*$ depends explicitly on the decay rate of the inflaton, $\Gamma_{\phi}$, and also implicitly,
since its value affects the effective equation of state during thermalization, 
as characterized by the {\it e-fold} average parameter $w_{\rm int}(\Gamma_{\phi})$ as well as 
the {\it time} average $w_{\rm eff}(\Gamma_{\phi})$ introduced previously.
In the previous section we have derived an estimate for $w_{\rm eff}$, finding that it has a universal value for $\Gamma_{\phi}\ll m$. The {\em e-fold} average of the equation of state $w_{\rm int}$ is given by
the time average of $w$ weighted by the Hubble parameter,
\beq
w_{\rm int} \equiv \frac{1}{N_{\rm reh}(\delta)-N_{\rm end}}\int_{N_{\rm end}}^{N_{\rm reh}(\delta)} w(n)\,dn
= \frac{1}{N_{\rm reh}(\delta)-N_{\rm end}}\int_{t_{\rm end}}^{t_{\rm reh}(\delta)} w(t)H(t)\,dt\,.
\eeq
Following the same procedure for $w_{\rm eff}$, we can approximate $w_{\rm int}$ as 
\begin{align}
w_{\rm int}^{(0)} &\approx \frac{1}{3\ln\left(\sqrt{\frac{3}{4}\rho_{\rm end}}\,t_{\rm reh}/M_P\right)}\int_0^{\Gamma_{\phi}t_{\rm reh}} \frac{ \boldsymbol{\gamma}({\textstyle \frac{5}{3}},u) }{ \boldsymbol{\gamma}({\textstyle \frac{5}{3}},u) +u^{2/3}e^{-u} }\,\frac{du}{u} \simeq \frac{0.731}{\ln(2.67m/\Gamma_{\phi})}\,,\\
w_{\rm int}^{(\infty)} &\approx \frac{0.743}{\ln(3.40m/\Gamma_{\phi})} \label{wint_est}\,,
\end{align}
for $\Gamma_{\phi}\ll m$ and $\delta=0.002$.

The previous semi-analytical results can be compared with the results of numerical integration of the
equations (\ref{eom1})-(\ref{eom3}). The dependence on the parameter $\delta$ of the effective 
{\it e-fold}-averaged equation-of-state parameter
$w_{\rm int}$ is shown in the right panel of Fig.~\ref{fig:wdelta}. 
Iteration from the first-order analytic approximation 
to the e-fold-averaged parameter $w_{\rm int}$ does not converge as rapidly as that for the time-averaged parameter $w_{\rm eff}$
(shown in the left panel of Fig. \ref{fig:wdelta} and in Fig.~\ref{fig:weff}). 
We see that the $\delta$-dependence of the iterated
approximation $w_{\rm int}^\infty$ for $\Gamma_\phi/m = 10^{-4}$(solid line) mirrors that of the numerical solution (red dots), 
though with a fractional offset $\lesssim 10$\%.

\begin{figure}[h!]
\centering
	\scalebox{0.8}{\includegraphics{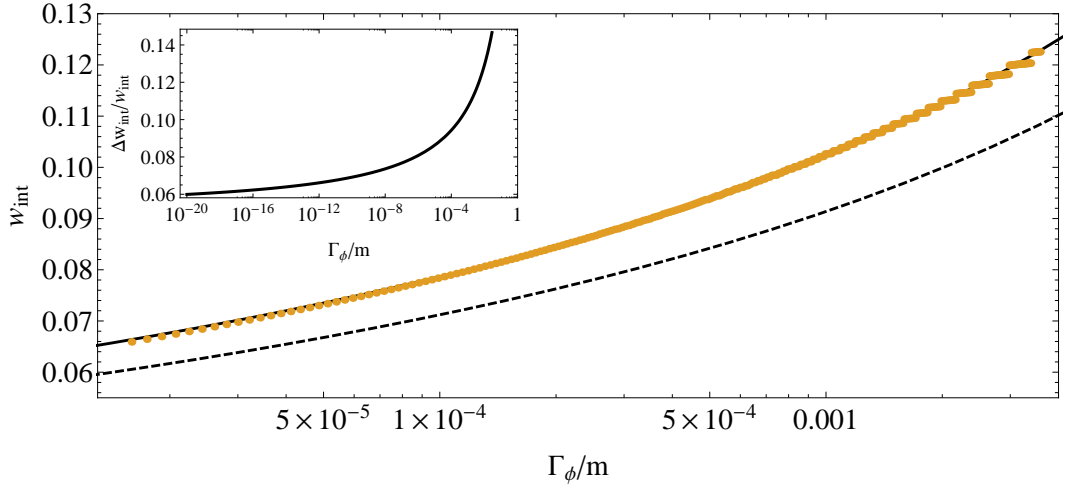}}
	\caption{\it The effective equation-of-state parameter $w_{\rm int}$ as a function of the decay rate of the inflaton. 
	The solid line corresponds to the fit (\protect\ref{w_fit}) to the data on the average of $w$, and 
	the dashed line represents the estimate (\protect\ref{wint_est}). 
	The inset displays the fractional difference between this approximate expression and the 
	fit (\protect\ref{w_fit}), which is small for $\Gamma_\phi \ll m$.}
	\label{fig:wint}
\end{figure}

Fig.~\ref{fig:wint} shows numerical values of $w_{\rm int}$ together with the estimate (\ref{wint_est}) as a function of $\Gamma_\phi/m$. Also displayed is a fit to the data, given by the equation
\beq\label{w_fit}
w_{\rm int} = \frac{0.782}{\ln(2.096m/\Gamma_{\phi})}\,.
\eeq
However, as already seen in the right panel of Fig.~\ref{fig:wdelta}, the fractional difference between the numerical result and the iterated analytic approximation
decreases as $\Gamma_\phi/m \to 0$, as seen in the insert in Fig.~\ref{fig:wint},
and is $\lesssim 10$\% for the range of $\Gamma_\phi/m$ of interest for our subsequent analysis.

\begin{figure}[h!]
\centering
	\scalebox{0.8}{\includegraphics{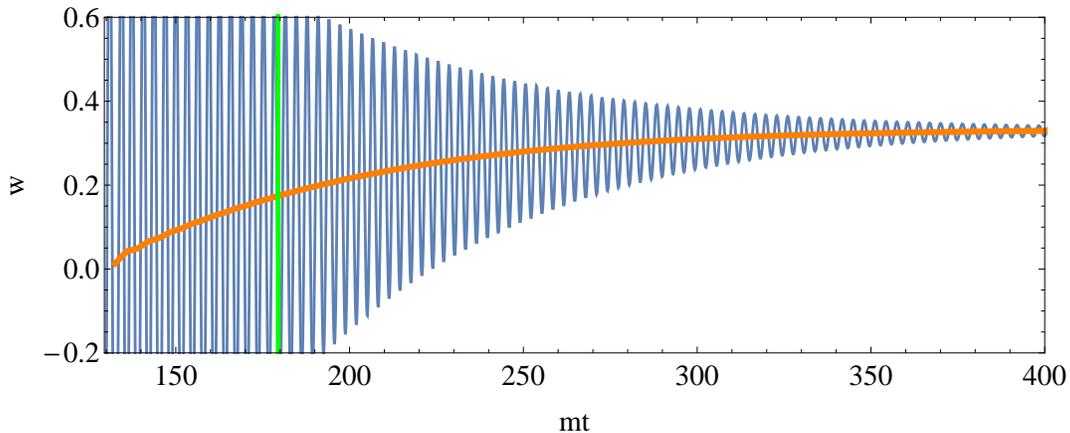}}
	\caption{\it The equation-of-state parameter $w(t)$ for a inflationary model with Starobinsky potential 
	and decay rate $\Gamma_{\phi}/m=2\times10^{-2}$. The inflaton field oscillations are shown as a blue line, and 
	the smooth (orange) curve interpolating between zero and 1/3 
	corresponds to the running average density ratio $\langle w \rangle$ (\protect\ref{womega}). 
	The vertical (green) line is located at the point $m (t_{\rm end} + 1/\Gamma_{\phi})$. } 
	\label{fig:womega}
\end{figure}

Fig.~\ref{fig:womega} illustrates the relationship between $w(t)$ and $\langle w \rangle$, 
which corresponds, by virtue of (\ref{womega}), to 1/3 of the energy density in radiation. As noted earlier, we
see explicitly that the estimate $t_{\rm reh}\sim \Gamma_{\phi}^{-1}$ (shown by the vertical green line) does not account fully for the decay of the inflaton into the 
relativistic degrees of freedom\footnote{Note that we have chosen a large value of $\Gamma_\phi/m$ to be able to 
see graphically the oscillations as a function of $mt$. For smaller $\Gamma_\phi/m$, the frequency of oscillations
would be larger and the details of the oscillations would disappear.}.

\begin{figure}[h!]
\centering
	\scalebox{0.8}{\includegraphics{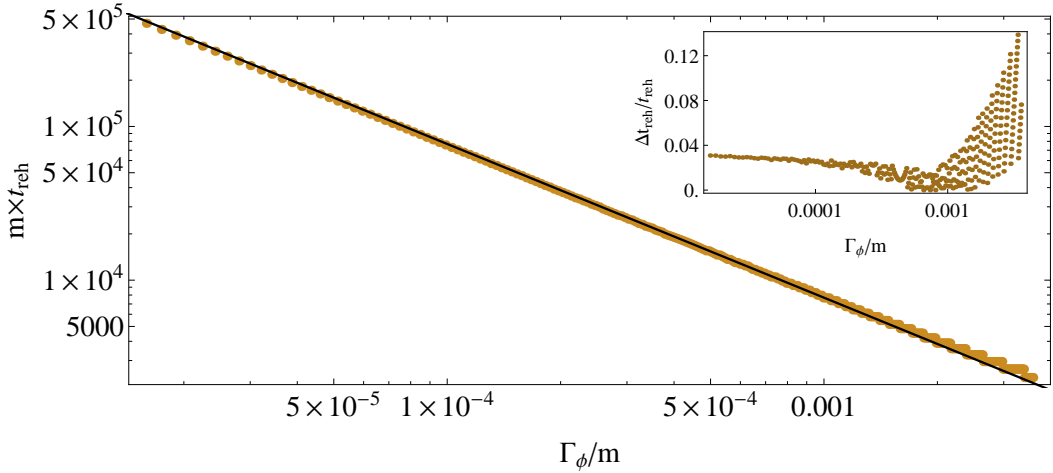}}
	\caption{\it A comparison between numerical and approximate analytical calculations of the end of reheating 
	as a function of the inflaton decay rate $\Gamma_{\phi}$. The (yellow) dots are obtained from the numerical integration
	of the equations of motion (\ref{eom1})-(\ref{eom3}). 
	The estimate (\ref{treh_app}) with $\delta=0.002$ is represented by the solid line. The inset displays the 
	fractional difference between the approximate expression $t_{\rm reh}^{(\infty)}$ and the exact numerical result.} 
	\label{fig:treh}
\end{figure}

\begin{figure}[h!]
\centering
	\scalebox{0.8}{\includegraphics{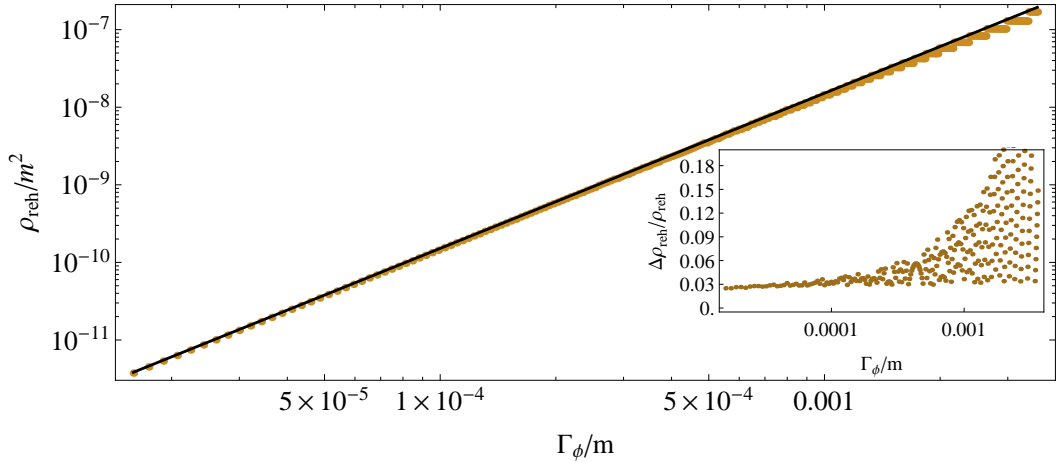}}
	\caption{\it Comparison between numerical and approximate analytical calculations of the energy density at the end of reheating 
	as a function of the inflaton decay rate $\Gamma_{\phi}$. The (yellow) dots are obtained by numerical integration
	of the equations of motion (\ref{eom1})-(\ref{eom3}). The solid line corresponds to the approximation (\ref{rhoreh}) with the value of $w_{\rm eff}$
	found by numerical integration. 
	The inset displays the difference between the approximate expression with $w_{\rm eff}\neq0$ and the exact numerical result. } 
	\label{fig:rhoreh}
\end{figure}

Fig.~\ref{fig:treh} shows the time at which reheating ends as a function of the decay rate. We see reasonable agreement between our 
analytical approximation (\ref{treh_app}) for $\delta = 0.002$ (solid line) and exact results found by numerical evaluation of the equations of motion 
(\ref{eom1})-(\ref{eom3}), which are represented by blue dots. 
The energy density at the end of reheating is displayed in Fig.~\ref{fig:rhoreh}, 
together with the approximation (\ref{rhoreh}) with the value of
$w_{\rm eff}$ that is determined by numerical integration. It is evident that the approximate expression is a good fit for the data, 
with a deviation $\lesssim 3\%$ for $\Gamma_{\phi}\ll m$, as shown in the insert in Fig.~\ref{fig:rhoreh}. 

\section{The Number of e-Folds in Representative No-Scale Inflation Models}

The preceding Section shows that we have good numerical and analytic control over the inflaton decay
and reheating process, which we now use to calculate the number of e-folds $N_*$ in some
representative no-scale models of inflation.

\begin{figure}[h!]
\vspace{1cm}
\centering
\hspace{-0.5cm}
	\scalebox{0.79}{\includegraphics{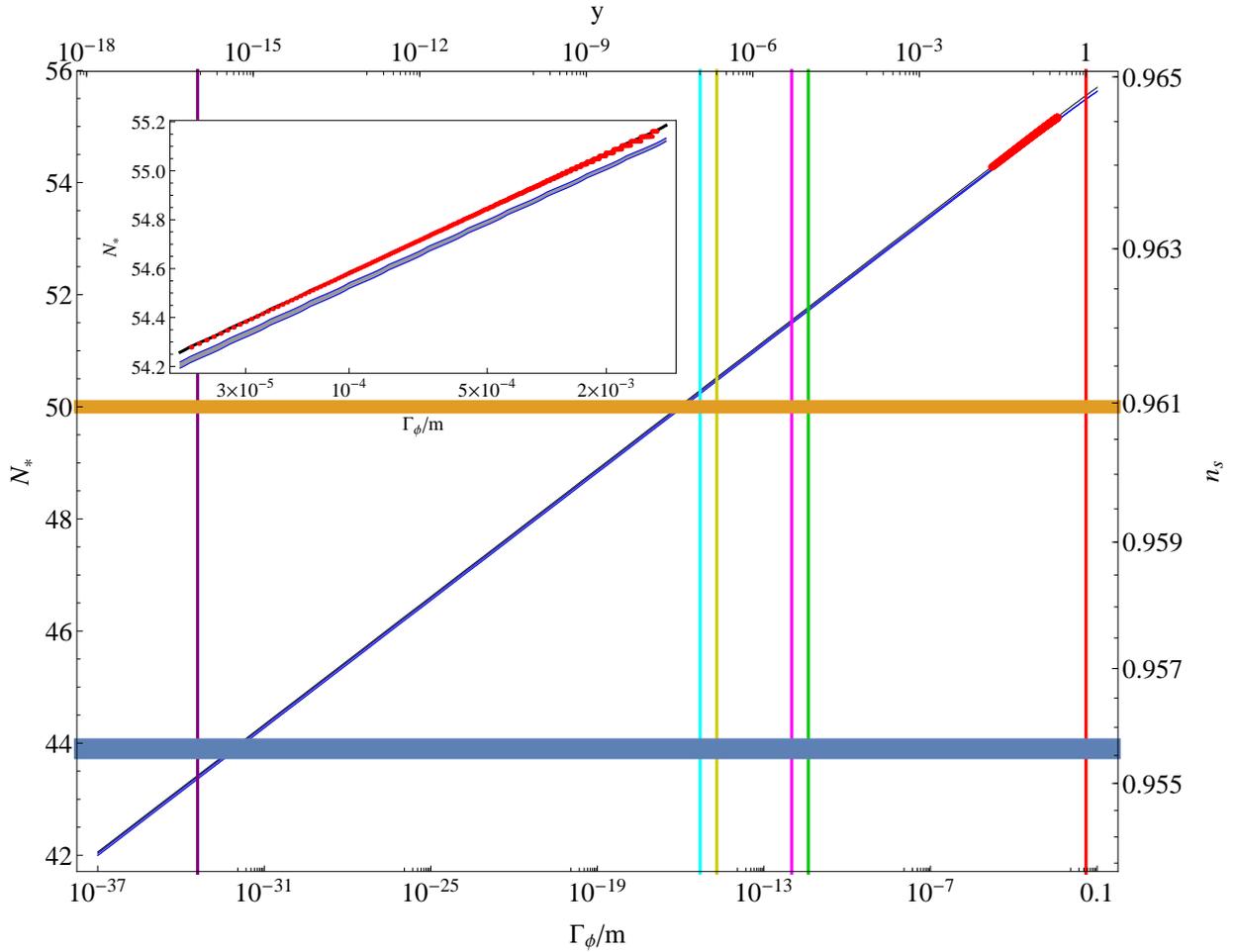}}
	\caption{\it The values of $N_*$ in no-scale Starobinsky-like models
	as a function of $\Gamma_\phi/m$, for a wide range of decay rates. The
	diagonal red line segment shows the full numerical results at $\delta=0.002$ over a restricted range of $\Gamma_\phi/m$, 
	which are shown in more detail in the insert, and the diagonal blue strip represents the analytical approximation 
	(\ref{NstarW}) for $10^{-3}<\delta<10^{-1}$. The difference between the results from evaluating $w_{\rm int}$ 
	via the iterative procedure and through the analytical approximation in the fit (\ref{w_fit}) are indistinguishable in the main plot, 
	but are visible in the insert, where the solid black line corresponds to (\ref{w_fit}). The right vertical axis shows
	the values of $n_s$ in Starobinsky-like no-scale models, for which
	the tensor-to-scalar ratio varies over the range $0.0034<r<0.0057$ for $N_*$ in the displayed range. The vertical
	coloured lines correspond to the specific models discussed in Section~3, and the horizontal yellow (blue) lines show the
	68 and 95\% CL lower limits from the Planck 2015 data, which vary slightly in related no-scale models,
	as discussed in Section~4.} 
	\label{fig:Nstar}
\end{figure}

We see from (\ref{res1}) that $N_*$ depends on $\Gamma_\phi$ both explicitly and implicitly via the dependences
in $w_{\rm int}$ and $w_{\rm eff}$, which have been shown in
(\ref{w_fit}) and Figs.~\ref{fig:weff} and \ref{fig:wint}. We use these in the general expression (\ref{res1})
to calculate $N_*$ as a function of $\Gamma_\phi$. For this purpose, we use the Planck pivot point
$k_* = 0.05$/Mpc, corresponding to $k_*/a_0 H_0 = 221$, and take the MSSM value of $g_{\rm reh} = 915/4$.
Fig.~\ref{fig:Nstar} displays the calculated value of $N_*$ over a wide range of $\Gamma_\phi$, parametrized by
\beq\label{rate_1}
\Gamma_{\phi} \; = \; m \frac{|y|^2}{8\pi}\,,
\eeq
with a coupling
ranging from
$y = 1$ 
to the value $y \simeq10^{-16}$, in which the latter would correspond to a reheating temperature $T_{\rm reh} \simeq 10$~MeV,
below which the successful conventional Big Bang nucleosynthesis calculations would need to be
modified substantially. Within this overall range, we discuss the values of $N_*$ found
in specific no-scale models whose inflaton decays were discussed in~\cite{EGNO4}.

\subsection{Decays via Superpotential Couplings}

In one class of model discussed in~\cite{EGNO4}, the inflaton was identified as
an untwisted matter field in some suitable string compactification, with direct decays to
matter particles via a Yukawa-like superpotential coupling. One possible realization of
this scenario would be further to identify the inflaton as a singlet (right-handed) sneutrino
$N$ with a couplings $y_\nu H L N$ to light Higgs and lepton doublets \cite{snu,ENO8}.
The perturbative decay rate of such a sneutrino inflaton would be given by
(\ref{rate_1}) with $y$ identified as the neutrino Yukawa coupling $y_\nu$. We use this as a
representative of the broader class of matter inflatons that decay directly to matter particles
via trilinear superpotential couplings.

Within the sneutrino inflation scenario, one might wish to consider values of $y_\nu \lesssim 1$,
the upper limit corresponding to a value of the Yukawa coupling similar to that of the top quark.
In this case, we estimate
\beq
y \; \simeq \; 1: \quad \quad N_* \; \simeq \; 55.5 \, ,
\label{y=1}
\eeq
as shown by the vertical red line in Fig.~\ref{fig:Nstar}. However, such a large value of $y$ would reheat the Universe to a very high temperature
$T_{\rm reh} \sim 10^{14}$~GeV, which would lead to an overproduction of gravitinos whose decays could aversely
affect big bang nucleosynthesis and could overpopulate the Universe with dark matter particles \cite{bbb}.
To avoid this overproduction, one should require $y_\nu \lesssim 10^{-5}$. 
For the upper limit in this case, we estimate
\beq
y \; \simeq \; 10^{-5}: \quad \quad N_* \; \simeq \; 51.7 \, ,
\label{y=10E-5}
\eeq
as shown by the vertical green line in Fig.~\ref{fig:Nstar}. On the other hand, as discussed above, the smallest value of $y$ consistent with conventional Big Bang nucleosynthesis
is $y \simeq 10^{-16}$, in which case
\beq
y \; \simeq \; 10^{-16}: \quad \quad N_* \; \simeq \; 43.4 \, ,
\label{y=10E-16}
\eeq
as shown by the vertical purple line in Fig.~\ref{fig:Nstar}. 
The above range of couplings includes possible gravitational decays of the inflaton~\footnote{As was discussed in~\cite{EGNO4}, another
possibility in such a matter inflaton scenario would be a superpotential coupling of the form $\zeta (T - 1/2)^2 \phi$,  which would yield
decays into $T$ fields with a rate
$\Gamma_{\phi} = m|\zeta|^2/(36\pi)$. The results for different values of $y$ discussed below
could also be applied to this case, by simply replacing $y \to \zeta \sqrt{2}/3$. }.
We discuss below the compatibility of these predictions with the Planck data,
shown as the horizontal yellow and blue lines in Fig.~\ref{fig:Nstar}.

\subsection{Decays via Gravitational-Strength Couplings}

There is another class of no-scale models in which the the inflaton decays via couplings
that are suppressed by one or more powers of $M_P$, which we exemplify here by examples in which the compactification volume
modulus $T$ is identified as the inflaton~\footnote{There are also matter-inflaton models in which decays are
suppressed by some power of $M_P$, but we do not discuss them here.}.
For instance, in the example discussed in Section~5.2 of~\cite{EGNO4},
there are decays into three-body $t {\bar t} H$ and related final states with rate
\beq\label{rate_2}
\Gamma(T\rightarrow H_u^0t_L\bar{t}_R, \, \tilde{t}_L\tilde{H}_u^0\bar{t}_R,\,\bar{\tilde{t}}_R t_L\tilde{H}_u^0)
\; = \; (2n_t+n_H-3)^2\frac{|y_t|^2m^3}{12(8\pi)^3M_P^2} \,,
\eeq
where $n_t$ and $n_H$ are modular weights that are ${\cal O}(1)$.
Since $y_t = {\cal O}(1)$ and $m \simeq 10^{-5} M_P$, this example corresponds to
\beq\label{gamma3}
{\rm 3-body~decay}: \quad \quad \frac{\Gamma_T}{m} \; \simeq \; 5 \times 10^{-16} \, .
\eeq
In this case we find
\beq
{\rm 3-body~decay}: \quad \quad N_* \; \simeq \; 50.3 \, ,
\label{3-body}
\eeq
as shown by the vertical pale blue line in Fig.~\ref{fig:Nstar}.

However, such three-body decays may be dominated by two-body inflaton decays into pairs of gauge bosons \cite{ekoty,klor},
if the gauge kinetic function $f_{\alpha \beta}$ (where $\alpha, \beta$ are gauge indices) has a
non-trivial dependence on the volume modulus: $f_{\alpha \beta} = f \delta_{\alpha \beta}$ with
\beq
d_{g, T} \; \equiv \; \langle {\rm Re} f \rangle^{-1} \left| \left\langle \frac{\partial f}{\partial T} \right\rangle \right| \; \ne \; 0 \, ,
\eeq
which is a generic feature of heterotic string 
effective field theories. In this case, the decays into Standard Model gauge bosons $V$ yield
\beq\label{rate_3}
\Gamma(T\rightarrow VV)
\; = \; \frac{d_{g, T}^2m^3}{32\pi M_P^2} \,,
\eeq
corresponding to
\beq\label{gammaVV}
{\rm Decays~into~gauge~bosons}: \quad \quad \frac{\Gamma_T}{m} \; \simeq \; \frac{d_{g, T}^2}{32\pi} m^2 \, .
\eeq
In a weakly-coupled heterotic string model, one might expect $d_{g, T} = {\cal O}(1/20)$,
whereas it might be ${\cal O}(1)$ in a strongly-coupled model, leading to
\begin{eqnarray} \label{weakstrong}
{\rm Weakly-coupled}: \quad \quad & \frac{\Gamma_T}{m} & \simeq \; 2 \times 10^{-15} \nonumber \\
{\rm Strongly-coupled}: \quad \quad & \frac{\Gamma_T}{m} & \simeq \;  10^{-12} \, .
\end{eqnarray}
These estimates of $\Gamma_\phi$ lead to the following estimates of $N_*$:
\begin{eqnarray} \label{weakstrongNstar}
{\rm Weakly-coupled}: \quad \quad N_* & \simeq & 50.5 \nonumber \\
{\rm Strongly-coupled}: \quad \quad N_* & \simeq &  51.5 \, ,
\end{eqnarray}
as shown by the vertical yellow and magenta lines in Fig.~\ref{fig:Nstar}, respectively.
The compatibility of these predictions with the Planck data is also discussed in the next Section.

\section{CMB Bounds on $\boldsymbol{N_*}$ in Representative No-Scale Inflation Models}

\subsection{Matter Inflaton Case}

In the recent no-scale inflation model~\cite{ENO6} with an untwisted matter field $\phi$ playing the role of the inflaton,
the observables $(n_s, r)$ were calculated assuming a K\"ahler potential of the form (\ref{noscaleK})
and choosing a Wess-Zumino superpotential $W(\phi)$ combining bilinear
and trilinear terms:
\beq
W \; = \; \frac{\mu}{2} \phi^2 - \frac{\lambda}{3} \phi^3 \, ,
\label{NSWZ}
\eeq
and assuming that the volume modulus $T$ is fixed. It was shown in~\cite{ENO6} that this model
reproduces exactly the predictions of the Starobinsky $R + R^2$ model if $\lambda = \mu/3$.
This model can alternatively be written in a more symmetric form:
\beq
K \; = \; - \, 3 \, \ln \left(1 - \frac{|y_1|^2 + |y_2|^2}{3} \right) \, ,
\label{symmK}
\eeq
where
\beq
y_1 \; = \; \left( \frac{2 \phi}{1 + 2 T} \right), \; y_2 \; = \; \sqrt{3} \left( \frac{1 - 2T}{1 + 2 T} \right) \, ,
\label{y1y2}
\eeq
in which representation the superpotential (\ref{NSWZ}) can be written as
\beq
W (y_1, y_2) \; = \mu \left[ \frac{y_1^2}{2} \left(1 + \frac{y_2}{\sqrt{3}} \right) - \frac{y_1^3}{3 \sqrt{3}} \right]  \, .
\label{Wy1y2}
\eeq
In the Starobinsky limit $\lambda = \mu/3$, and we consider related models with $\lambda/\mu  \sim 1/3$.

\begin{figure}[h!]
\centering
\hspace{0.5cm}
\scalebox{0.7}{\includegraphics{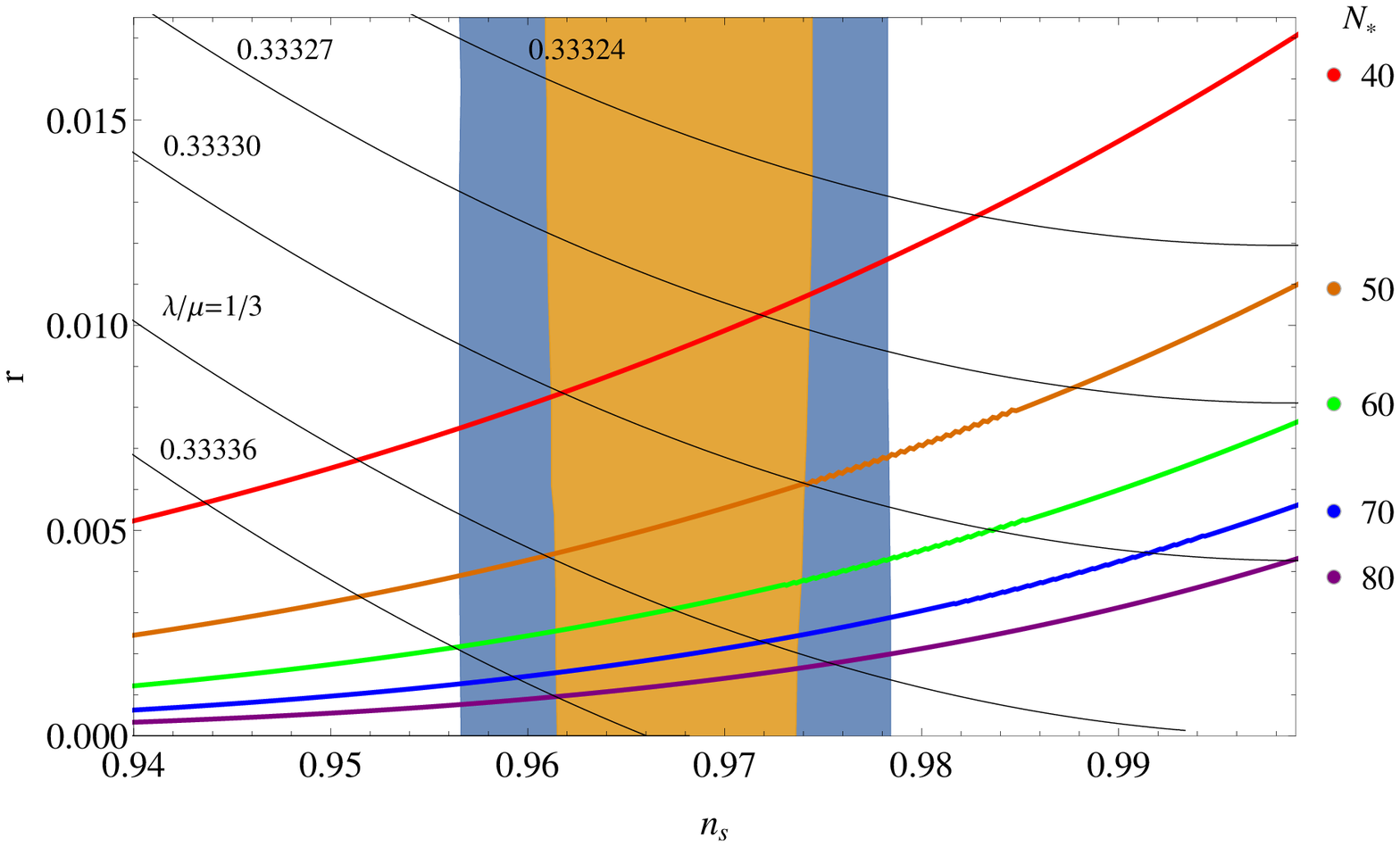}} \\
\hspace{-0.7cm}
\scalebox{0.73}{\includegraphics{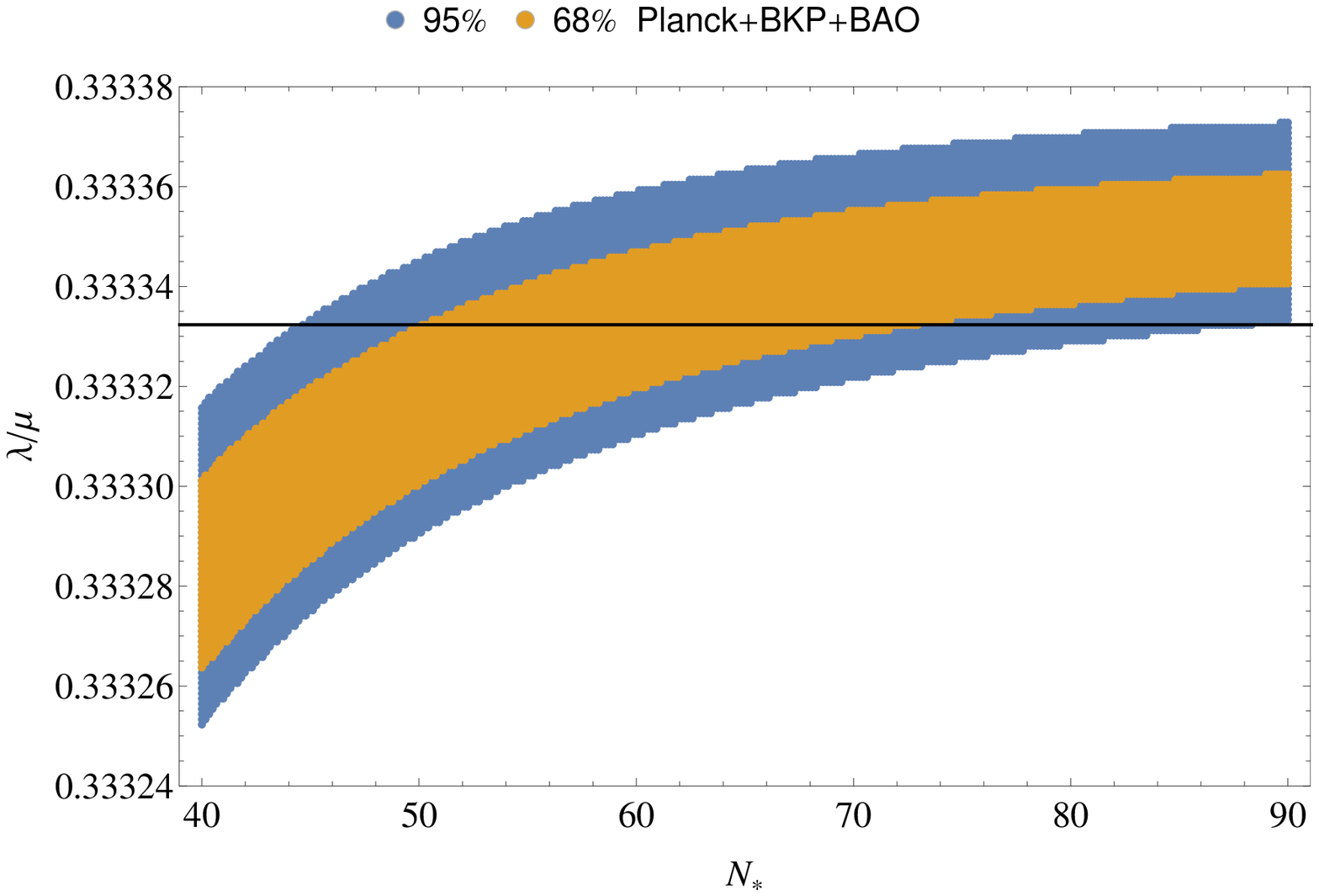}}
	\caption{\it The 68\% and 95\% CL regions (yellow and blue, respectively) in the $(n_s, r)$ plane (upper panel)
	and the $(N_*, \lambda/\mu)$ plane (lower panel)
	for the no-scale inflationary model~\protect\cite{ENO6} with a matter inflaton field and the Wess-Zumino superpotential
	(\protect\ref{NSWZ}). The black lines in the upper panel are contours of $\lambda/\mu$,
	and the coloured lines are contours of $N_*$.
	The horizontal black line in the lower panel is for $\lambda/\mu = 1/3$, the value that reproduces the
	inflationary predictions of the Starobinsky model~\protect\cite{ENO6}.} 
	\label{fig:NSWZ}
\end{figure}

The calculations of $N_*$ in the previous Sections were made assuming exactly Starobinsky-like
inflation, which (as already mentioned) corresponds in this matter inflation model to the limiting case $\lambda/\mu = 1/3$.
We have studied the modification of the $N_*$ calculation when $\lambda/\mu \ne 1/3$, but lying
within the range $0.33324 \le \lambda/\mu \le 0.33338$ displayed in Fig.~\ref{fig:NSWZ}.

Fig.~\ref{fig:NSWZ} displays the Planck 2015 constraints on this model in the $(n_s, r)$
plane (upper panel) and the $(N_*, \lambda/\mu)$ plane (lower panel),
with the region favoured at the 68\% CL shaded yellow, and the region allowed at the 95\% CL
shaded blue. We see in the upper panel that for values of $\lambda/\mu \sim 1/3$
(black lines) the tensor to scalar ratio is small, and in this case, 
the data yield constraints on $n_s$ that are relatively insensitive to $r$.
On the other hand, we see that any fixed value of $n_s$ corresponds to values of $N_*$
(coloured lines) that are strongly correlated with the values of $\lambda/\mu$. Thus for 
a given value of $\lambda/\mu$, the lower bound on $n_s$ provided by Planck can be translated into
a lower bound on $N_*$ that is sensitive to $\lambda/\mu$. For example, for $\lambda/\mu = 1/3$, the 68\% lower bound on $n_s$ 
corresponds to a lower bound of $\simeq 50$ on $N_*$. 

This feature is reflected in the lower panel of Fig.~\ref{fig:NSWZ},
where we see that $N_*$ is essentially unconstrained in this model in the absence of a
precise value for $\lambda/\mu$. However, if one assumes the Starobinsky value $\lambda/\mu = 1/3$,
one finds $N_* \in (50, 74)$ at the 68\% CL, which would disfavour $y \lesssim 10^{-9}$
according to Fig.~\ref{fig:Nstar}, and the 68\% CL
lower bound on $N_*$ would strengthen for $\lambda/\mu > 1/3$.

\begin{figure}[h!]
\centering
\scalebox{0.85}{\includegraphics{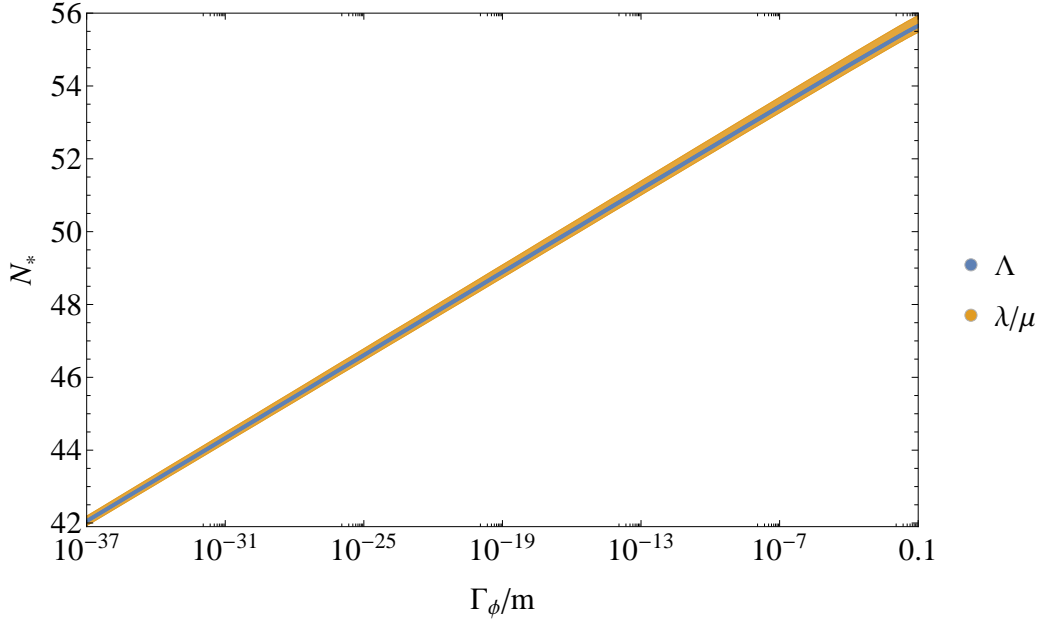}}
	\caption{\it The possible variation of the value of $N_*$
	around the prediction for the Starobinsky limit $\lambda/\mu = 1/3$ in the matter inflation
	model (\protect\ref{NSWZ}) with $0.33324 \le \lambda/\mu \le 0.33338$ (yellow band),
	and for a stabilizing parameter $10^{-2} \le \Lambda \le 1$  in the \kahler potential (\ref{quarticK}) (blue band), 
	as a function of the inflaton decay rate $\Gamma_\phi$.} 
	\label{fig:Nlambdamu}
\end{figure}

 As seen in
Fig.~\ref{fig:Nlambdamu}, the maximum deviation from the Starobinsky prediction for $N_*$
(as shown in Fig.~\ref{fig:Nstar})
due to varying $\lambda/\mu$ in the range studied (yellow band) is always
$\lesssim 1$, and the deviation is significantly smaller for the favoured models with inflaton decay via a
two-body superpotential coupling $y \lesssim 10^{-5}$ (corresponding to the green vertical line in Fig.~\ref{fig:Nstar}).
The Starobinsky-like analysis gave $N_* \simeq 51.7$ for $y = 10^{-5}$, as seen in (\ref{y=10E-5}), and
the non-Starobinsky deviation of $N_*$ in Fig.~\ref{fig:Nlambdamu} is $\lesssim 0.5$ for this value of $y$,
decreasing to much smaller values close to the Big Bang nucleosynthesis lower limit $y \simeq 10^{-16}$,
for which we found $N_* \simeq 43.4$ in the Starobinsky limit, as seen in (\ref{y=10E-16}).

It is necessary to address in this model two potential issues: the stabilization of the real and imaginary parts of the field $y_2$
when the inflaton $y_1$ reaches its minimum, and the possibility that ${\cal R}e\, y_2 \ne 0$ during inflation.
The first of these issues is resolved by adding a small supplementary term to the superpotential (\ref{Wy1y2}):
\beq
\Delta W \; = \; b \, \mu \, \frac{y_2^2}{3} \, ,
\label{DeltaW}
\eeq
for some constant $b$.
The second issue is addressed by incorporating a quartic term
in the K\"ahler potential (\ref{symmK}):
\beq
K \; = \; - \, 3 \, \ln \left(1 - \frac{|y_1|^2 + |y_2|^2}{3} + \frac{|y_2|^4}{\Lambda^2} \right) \, ,
\label{quarticK}
\eeq
where typical values of $\Lambda \lesssim 1$ in natural units, as discussed in~\cite{ENO7}.

\begin{figure}[h!]
\centering
\scalebox{0.7}{\includegraphics{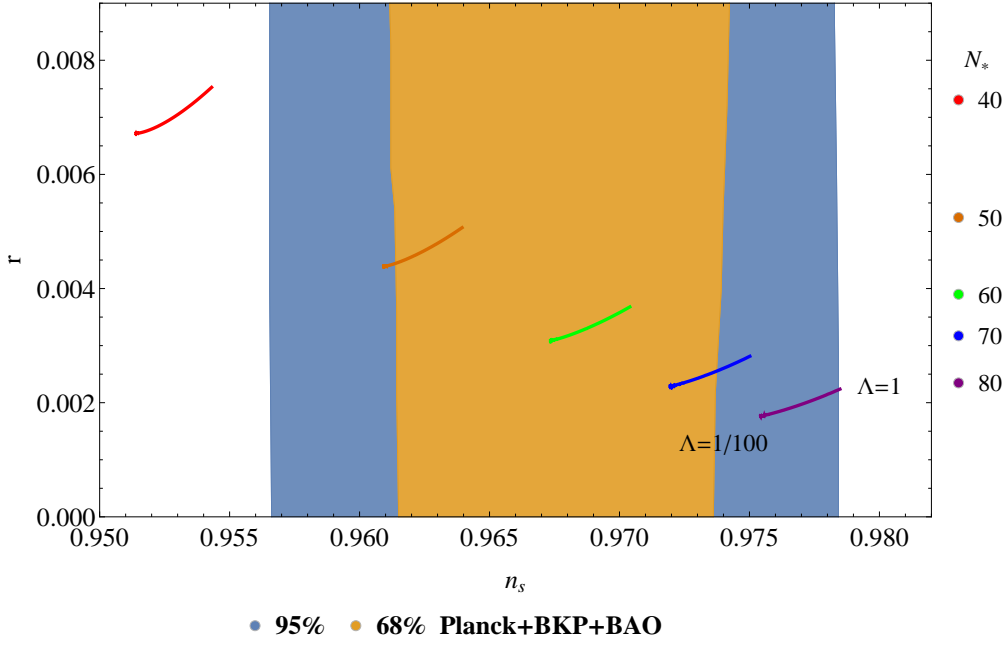}} \\
\hspace{-0.7cm}
\scalebox{0.7}{\includegraphics{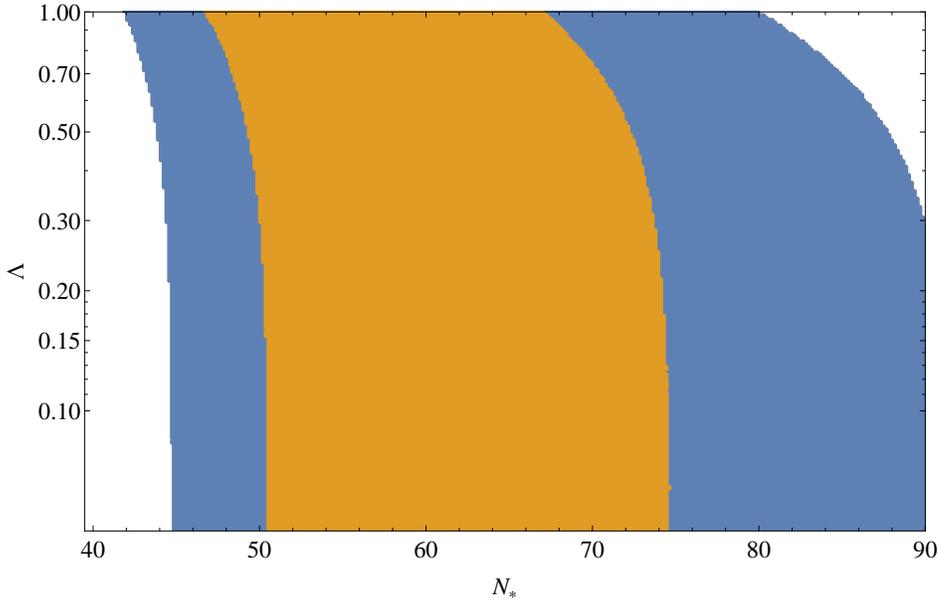}}
	\caption{\it The 68\% and 95\% CL regions (yellow and blue, respectively) in the $(n_s, r)$ plane (upper panel)
	and the $(N_*, \Lambda)$ plane (lower panel)
	for the no-scale inflationary model~\protect\cite{ENO7} with a matter inflaton field and the Wess-Zumino superpotential
	(\protect\ref{Wy1y2}, \protect\ref{DeltaW}), for $b = 10^{-6}$ and $\Lambda \le 1$.
	The coloured lines in the upper panel are contours of $N_*$ for $0.01 \le \Lambda \le 1$.} 
	\label{fig:Lambda}
\end{figure}

Fig.~\ref{fig:Lambda} shows the 68\% and 95\% CL regions (yellow and blue, respectively)
in the $(n_s, r)$ plane (upper panel) and in the $(N_*, \Lambda)$ plane (lower panel)
for this no-scale inflationary model with the illustrative choice $b = 10^{-6}$ and the range $\Lambda \le 1$. 
As shown in~\cite{ENO7},
this matter inflaton model reproduces the inflationary predictions of the Starobinsky model for $\Lambda \lesssim 1$,
and we see in the upper panel of Fig.~\ref{fig:Lambda} that the data constraints on $n_s$ are
insensitive to $r$ for $\Lambda$ in this range and the relevant values of $N_*$ (coloured lines). 
In each of the segments shown, $\Lambda$ varies from 1/100 to 1 as shown for several values of $N_*$.
Once again, we can use the lower bound on $n_s$ to derive a lower bound on $N_*$ for a 
given value of $\Lambda$. Those limits are reproduced 
in the lower panel of Fig.~\ref{fig:Lambda} where we see that in the limit of small $\Lambda$ 
the current data require $N_* > 45.5$ at the 95\% CL and favour
$N_* \in (51.0, 75.2)$ at the 68\% CL. The variation in $N_*$ for values of $\Lambda \in (10^{-2}, 1)$
is shown as a thin blue band in Fig.~\ref{fig:Nlambdamu} as a function of $\Gamma_\phi/m$.
We see that this is always smaller than the variation due to varying $\lambda/\mu$,
being $\ll 1$ and hence negligible for our purposes.

\subsection{Volume Modulus Inflaton Cases}

In~\cite{EGNO2, EGNO3} we calculated the observables $(n_s, r)$ in various
no-scale inflationary models in which the inflaton was identified with some combination of the real
and imaginary parts of the complex volume modulus $T$. 
In this case, we choose
a superpotential of the form \cite{Cecotti}
\beq\label{tph_w}
W=\sqrt{3}m\phi(T-1/2) \, .
\eeq
Inflation along the direction of the
canonically normalized real component $\rho$ of $T$ yielded a Starobinsky-like model, whereas there was
a quadratic potential along the imaginary direction \cite{FeKR,EGNO}. In~\cite{EGNO2} we assumed that higher-order
terms in the K\"ahler potential $K$ fixed the angle of the inflationary trajectory in the complex $T$ plane,
whereas~\cite{EGNO3} we made a complete two-field analysis of the inflationary observables $n_s$ and $r$.
We now confront these models with the Planck 2015 data.

The models are characterized by two parameters, the angle of the starting-point in the complex $T$
plane, which we parameterize here as $\alpha \equiv \arctan (\sigma/\rho)$~\footnote{Note that the angle
$\theta$ used in~\cite{EGNO2, EGNO3} is equivalent to $\pi/2 - \alpha$.}, and the modulus stabilization
parameter $c$~\cite{EKN}:
\begin{equation}
K \; = \; -3\ln\left(T+T^* + c\left[ \sin \alpha(T+T^*-1) - \cos \alpha(T-T^*)^2 \right]^2 \right) + \frac{|\phi|^2}{(T+T^*)^3} \, .
\label{stabilize}
\end{equation}
In this model the matter field $\phi$ relaxes dynamically to zero during inflation~\cite{EGNO2}. As was discussed in~\cite{EGNO2, EGNO3}, the case $\alpha = \pi/2$ corresponds to a quadratic model of chaotic
inflation, and smaller values of $\alpha$ interpolate between this and the Starobinsky limit when $\alpha = 0$.
If the quartic stabilization term $\propto c$ is large enough,
the inflaton trajectory follows a narrow valley in field space, much like a bobsleigh run, whereas if $c$ is
small the inflaton trajectory is less constrained and two-field effects become important. In this case, we found in~\cite{EGNO3}
that the trajectories for $\alpha < \pi/2$ tend to become more Starobinsky-like for smaller values of $c$.
Predictions of this model for various values of $c$ and $\alpha$, based on a full two-field analysis,
can be found in~\cite{EGNO3}, where it can be seen that the results depend on $N_*$. The Planck 2015
68\% and 95\% CL contours in the $(n_s, r)$ plane can then be converted into the corresponding
constraints on $N_*$ in these no-scale models as functions of $c$ and $\alpha$, as illustrated in the
following figures.

In Fig.~\ref{fig:c100} we display the 68\% and 95\% CL regions (yellow and blue, respectively) in the $(n_s, r)$ plane
(upper panel) and in the $(N_*, \alpha)$ plane (lower panel) for the
strongly-stabilized case $c = 100$~\footnote{Larger values of $c$ would give very similar results.}.
We see in the upper panel that in this case the data constraints on $n_s$ depend in an essential way
on the value of $r$, and we also see that the contours of $N_*$ (coloured lines) depend in a non-trivial way on the value
of $\alpha$, as discussed in~\cite{EGNO3}. The interpolation between $\alpha = \pi/2$, the limit in which the model
realizes chaotic quadratic inflation, and $\alpha = 0$, the Starobinsky limit, is nonlinear.

\begin{figure}[h!]
\centering
\hspace{0.5cm}
\scalebox{0.7}{\includegraphics{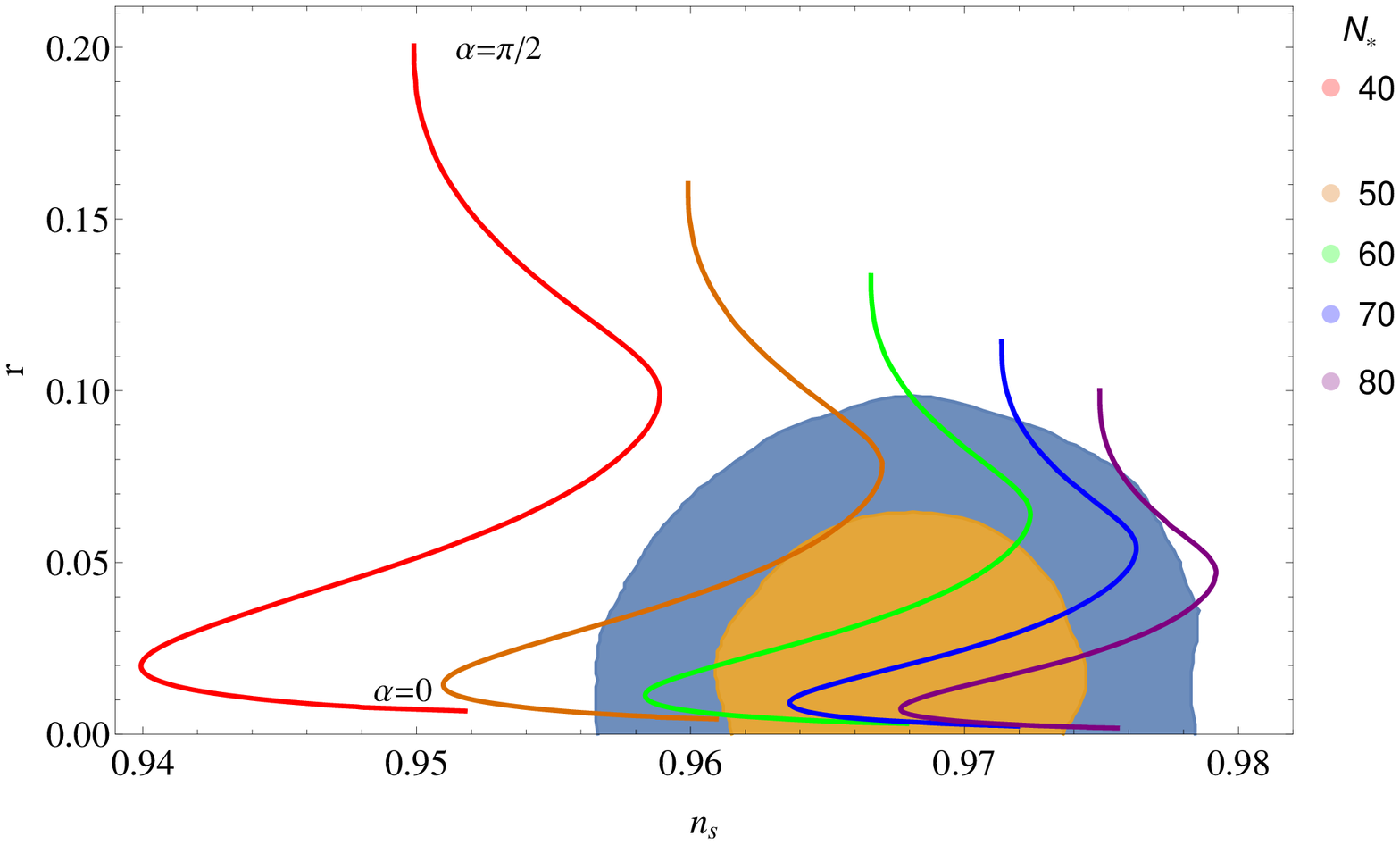}} \\
\hspace{-0.5cm}
\scalebox{0.75}{\includegraphics{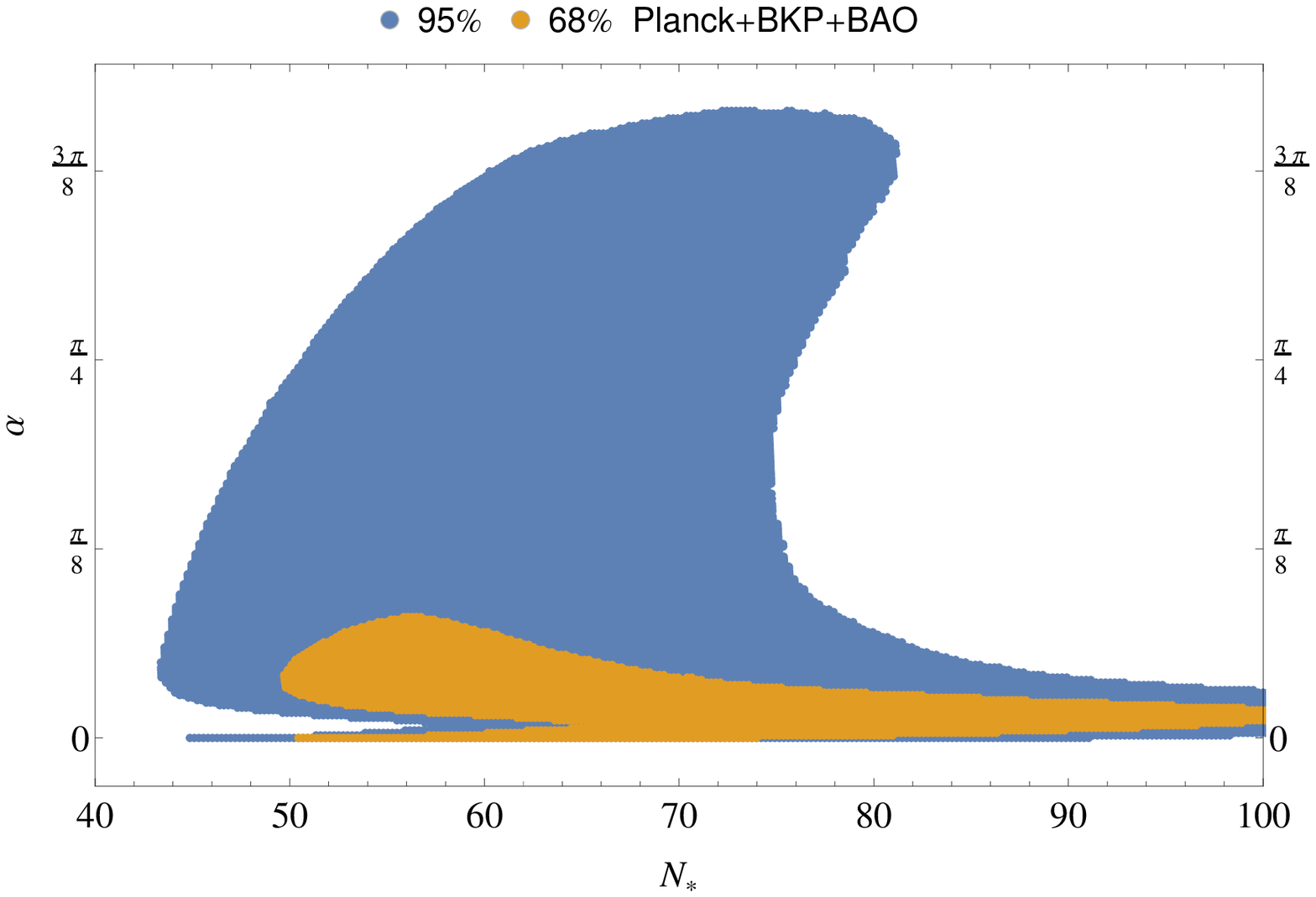}}
	\caption{\it The 68\% and 95\% CL regions (yellow and blue, respectively) in the $(n_s, r)$ plane (upper panel)
	and the $(N_*, \alpha)$ plane (lower panel)
	for the no-scale inflationary model (\protect\ref{stabilize}), assuming strong stabilization with $c = 100$.
	The coloured lines in the upper panel are contours of $N_*$ for $0 \le \alpha \le \pi/2$.} 
	\label{fig:c100}
\end{figure} 

\begin{figure}[h!]
\centering
\scalebox{0.85}{\includegraphics{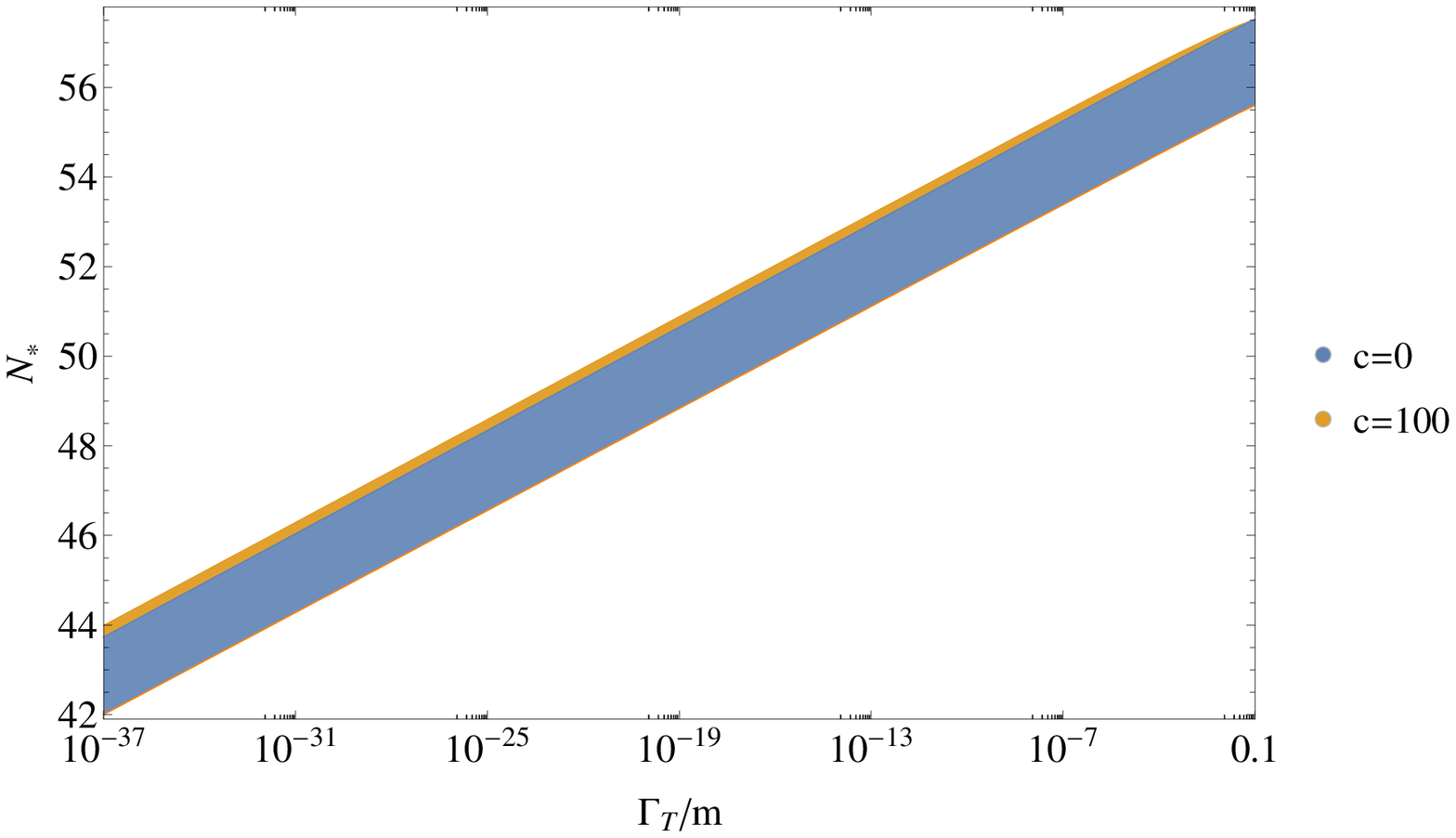}}
	\caption{\it The variation of the value of $N_*$ as a function of the normalized
	decay rate $\Gamma_T/m$ for the modulus inflaton model (\protect\ref{stabilize}) with $c=100$ and $c=0$
	(blue and yellow bands, respectively), in the range $0\le \alpha\le \pi/2$. The stabilized region $c=100$ contains fully the
	unstabilized region with $c=0$. The upper sides of the bands correspond to the limit of chaotic inflation with a
	quadratic potential, and the lower sides to the Starobinsky limit.} 
	\label{fig:NstarT}
\end{figure}

For these reasons, the conversion of the CMB data into constraints on $N_*$ also depends
non-trivially on $\alpha$, as seen in the ``whale-like'' shape in the lower panel of Fig.~\ref{fig:c100}.
In particular, the whale's ``mouth" is the converse of the leftward swerve in the $N_*$ contours in
the upper panel of Fig.~\ref{fig:c100}, and the ``lower jaw" corresponds to the swerve back to larger $n_s$
at small $r$.
Overall, we see that $N_* \ge 50$ is preferred at the 68\% CL, whereas $N_* \ge 43$ is allowed at the 95\% CL.
Translating these limits into constraints on $\Gamma_T/m$ is also less trivial than in the previous
matter inflaton models, since this matter inflaton model is not Starobinsky-like for large $\alpha$, as
seen in the upper panel of Fig.~\ref{fig:c100}. As seen in Fig.~\ref{fig:NstarT}, any fixed value of $N_*$
may correspond to a range of values of $\Gamma_T/m$, shown by the blue band, depending on the value of $\alpha$.
The upper side of the band correspond to the limit of chaotic inflation with a quadratic potential, 
and the lower side to the Starobinsky limit. As we see in Fig.~\ref{fig:Nstar}, in the Starobinsky-like
limit $\alpha \to 0$ we find the following constraints on $\Gamma_T/m$ and the effective two-body coupling $y$:
\beq
\frac{\Gamma}{m} \; \gtrsim \; 3 \times 10^{-20}, \; \; y \; \gtrsim \; 10^{-9} \; \; (68\% \; {\rm CL}) \, .
\eeq
We have studied numerically the possible variations in the dependence of $N_*$
on the decay rate $\Gamma_T$ in the range $0\le \alpha\le \pi/2$,
with the results shown as the blue band in Fig.~\ref{fig:NstarT}. 
We see there that in the chaotic quadratic inflation case the constraint
on $\Gamma_T/m$ is relaxed by a factor $\sim 10^6$ and the effective two-body coupling $y$
is relaxed correspondingly by a factor $\sim 10^3$.

If $y \lesssim 10^{-5}$, 
corresponding to values of $N_* \lesssim 52.7$ as shown in (\ref{y=10E-5})
and Fig.~\ref{fig:Nstar} for the Starobinsky case, then only the range $\alpha \lesssim \pi/16$ is
allowed at the 68\% CL, rising to $\alpha \lesssim \pi/4$ at the 95\% CL.
Thus models with a starting-point of inflation with small $\alpha$ close to the real direction, i.e., 
close to the Starobinsky model, are favoured in the strongly-stabilized case, though not strongly at the 95\% CL.
In the Starobinsky limit $\alpha \to 0$, corresponding to the lower jaw of the ``whale" in Fig.~\ref{fig:c100},
the value $y \sim 10^{-5}$ is at the Planck 2015 68\% CL limit, whereas $N_* \gtrsim 47$
and hence $y \gtrsim 10^{-13}$ is allowed at the 95\% CL.

\begin{figure}[h!]
\centering
\scalebox{0.7}{\includegraphics{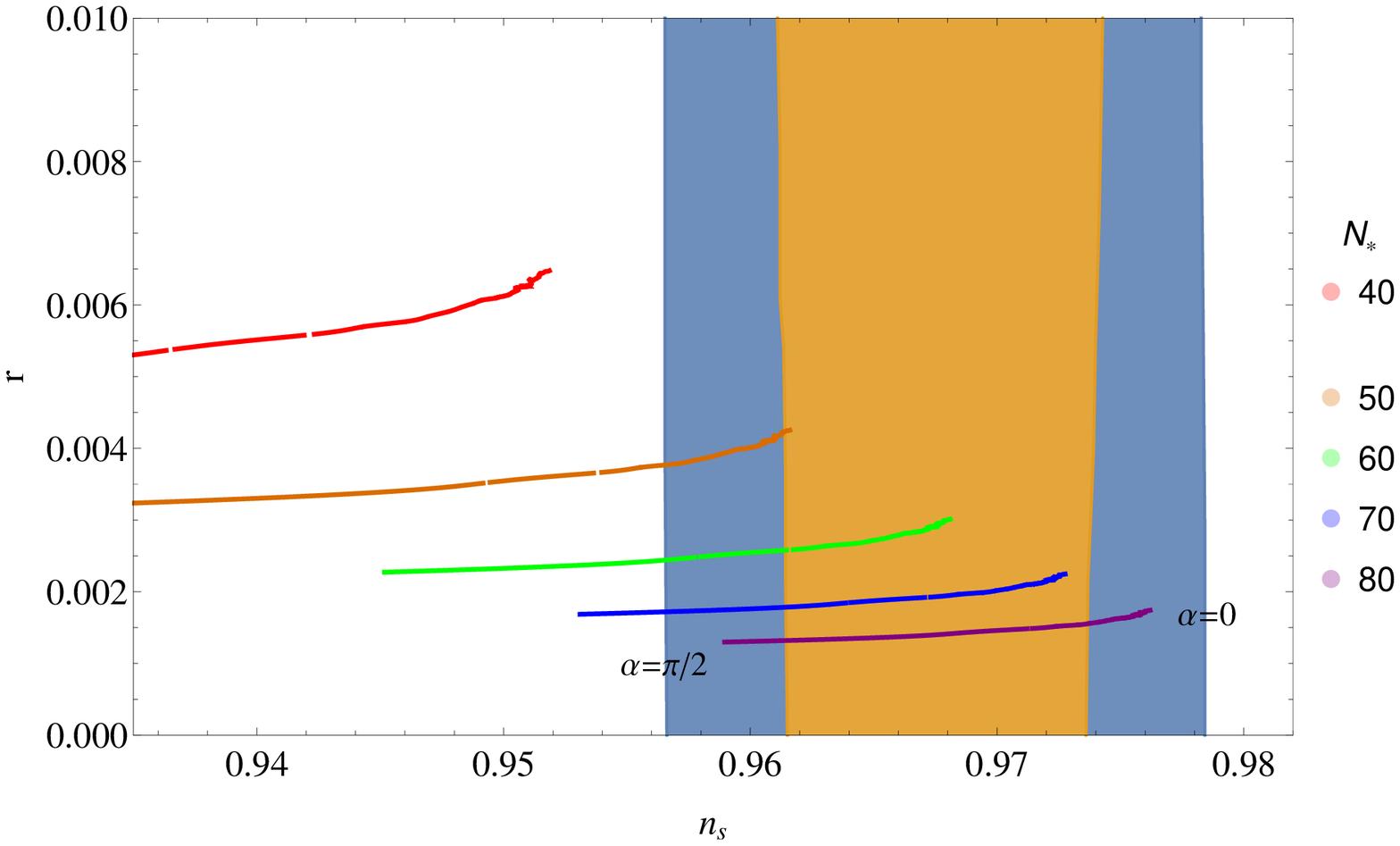}} \\
\hspace{-0.4cm}
\scalebox{0.7}{\includegraphics{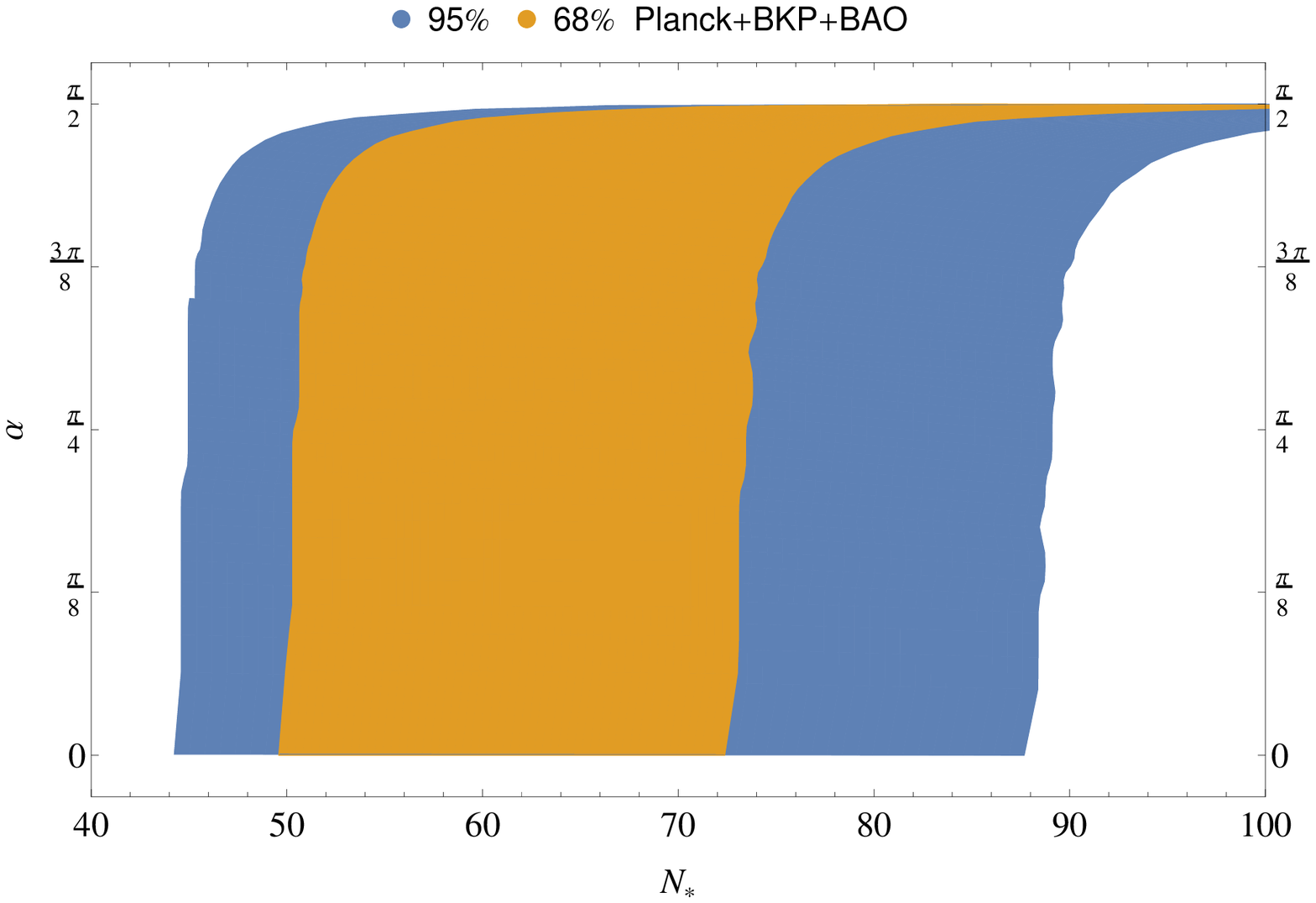}}
	\caption{\it As for Fig.~\protect\ref{fig:c100}, but assuming no stabilization, i.e., $c = 0$.} 
	\label{fig:c0}
\end{figure}

Fig.~\ref{fig:c0} shows analogous results for the unstabilized case $c = 0$. In this case,
as seen in the upper panel, the important CMB constraint is that from $n_s$, which is
almost independent of $r$ for the relevant values of $N_*$ (coloured lines).
The lower panel of Fig.~\ref{fig:c0} shows that again
$N_* \ge 50$ is favoured at the 68\% CL, whereas now $N_* \ge 44$ is allowed at the 95\% CL.
Both of these limits are rather insensitive to $\alpha$ in the range $\lesssim 3 \pi/8$, whereas only
very large values of $N_*$ are allowed as $\alpha \to \pi/2$, corresponding to a starting-point
along the imaginary direction. In this case there is no significant preference for Starobinsky-like
models with the starting-point of inflation close to the real direction.

\begin{figure}[h!]
\centering
\scalebox{0.65}{\includegraphics{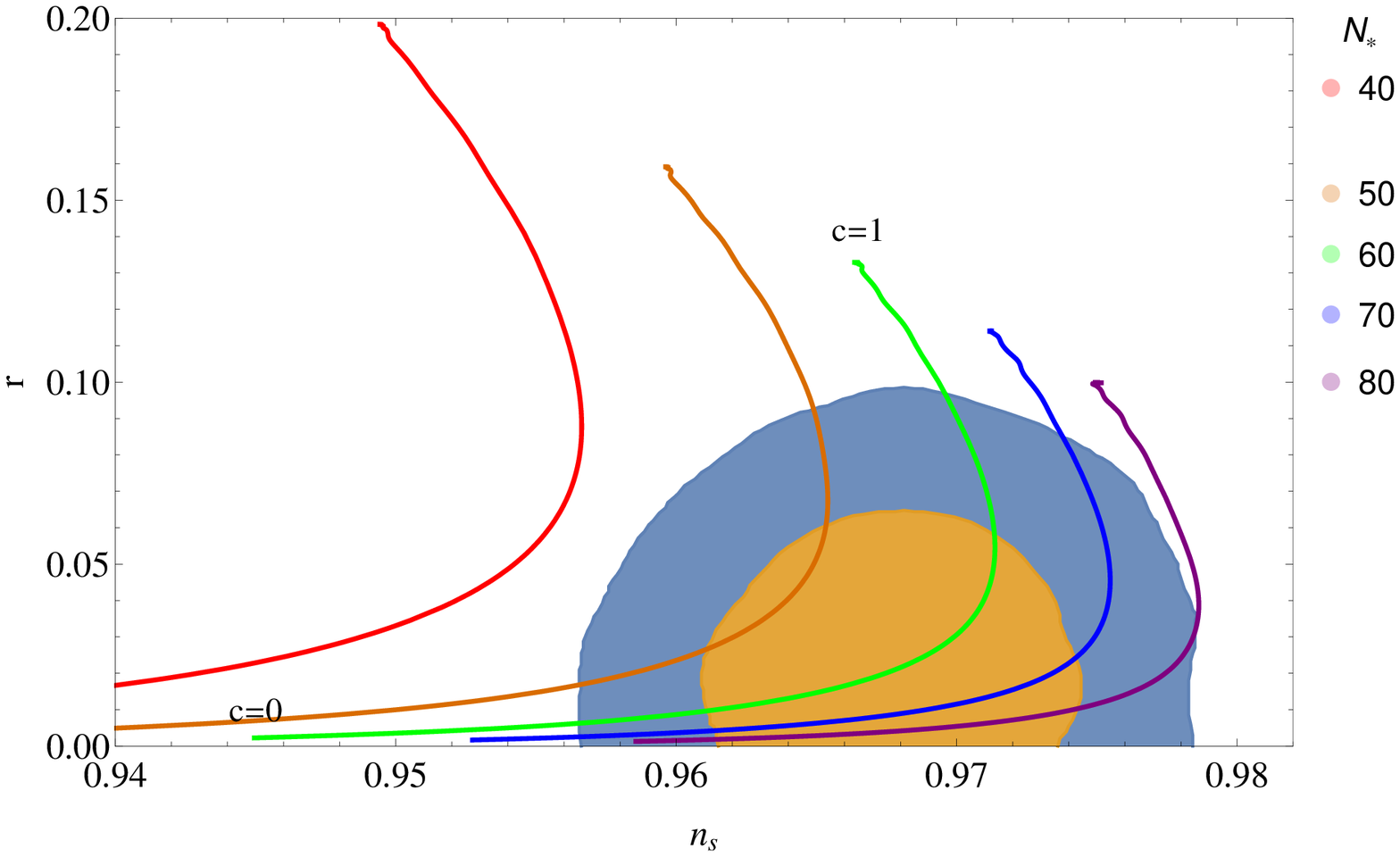}} \\
\hspace{-0.6cm}
\scalebox{0.7}{\includegraphics{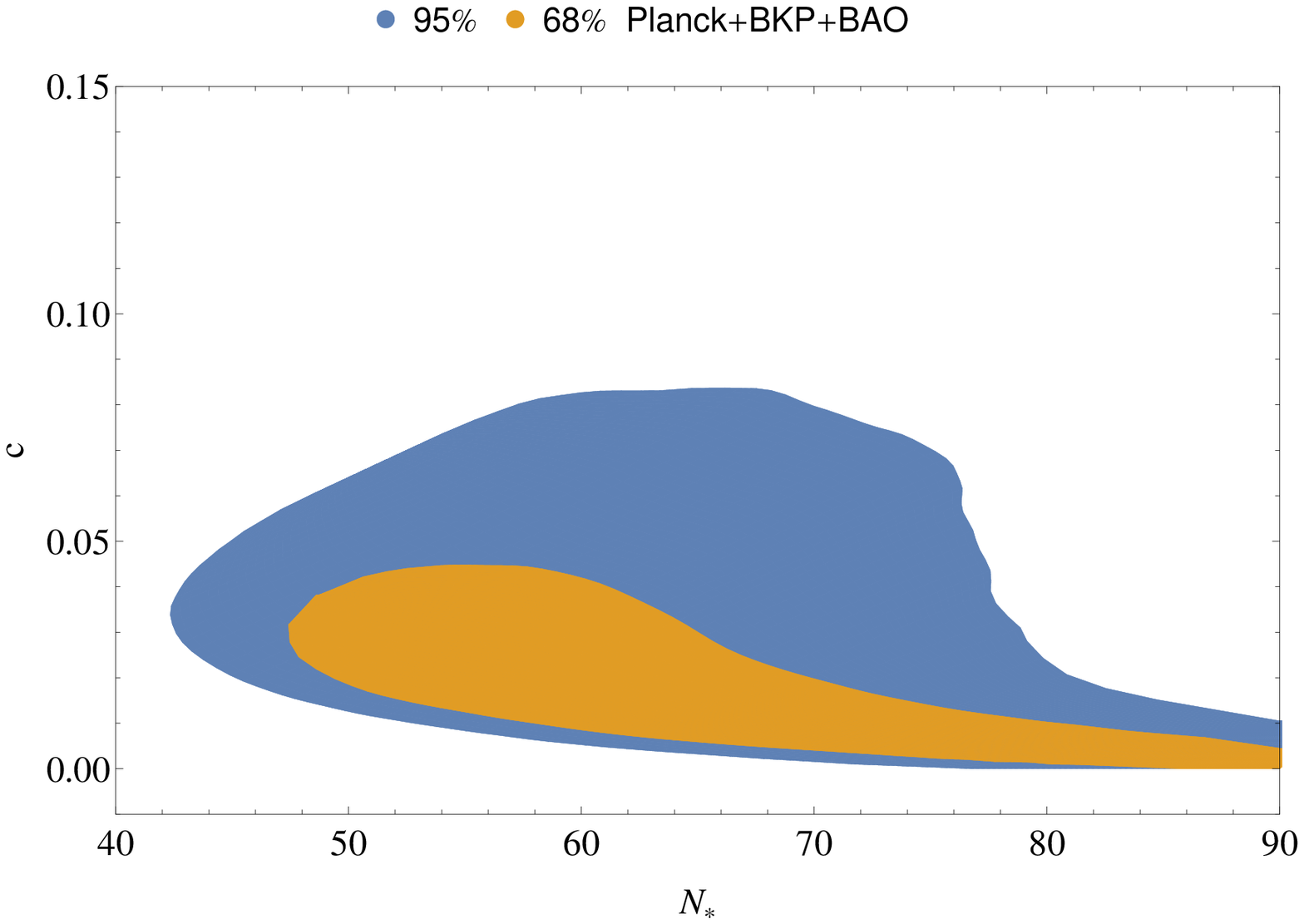}}
	\caption{\it The 68\% and 95\% CL regions (yellow and blue, respectively) in the $(n_s, r)$ plane (upper panel)
	and the $(N_*, c)$ plane (lower panel)
	for the no-scale inflationary model (\protect\ref{stabilize}), assuming a starting-point along the imaginary
	field direction. The coloured lines in the upper panel are contours of $N_*$ for $0 \le c \le 1$.} 

	\label{fig:im}
\end{figure}

Fig.~\ref{fig:im} shows analogous results for inflation along the imaginary direction, i.e., $\alpha = \pi/2$,
and allowing the stabilization parameter $c$ to vary. As we see in the upper panel, the cosmological
constraint depends non-trivially on both $n_s$ and $r$, and the values of these quantities for fixed $N_*$
(coloured lines) depend non-linearly on $c \in [0, 1]$. In this case, as seen in the lower panel of Fig.~\ref{fig:im},
the lowest value of $N_*$ is found for $c \simeq 0.03$, with
$N_* \ge 48$ favoured at the 68\% CL and $N_* \ge 43$ allowed at the 95\% CL.
As can be seen in Fig.~\ref{fig:NstarT}, these constraints correspond to $y \gtrsim 10^{-12}$
being favoured at the 68\% CL, whereas all values of $y$ consistent with Big Bang nucleosynthesis are allowed at
the 95\% CL. This example shows that inflation is possible in this model even if the starting-point of
inflation is close to the imaginary direction, far from the Starobinsky-like limit.

\section{Summary and Prospects}

The first purpose of this paper has been to present in Section~2 calculations of $N_*$ in models of
inflaton decays, with particular attention to predictions in Section~3 from models that are motivated by no-scale models
of inflation and have a structure inspired by common scenarios for string compactification.
We have then analyzed in Section~4 the ranges of $N_*$ within no-scale models that yield values of
$n_s$ and $r$ within the 68 and 95\% CL regions found by the Planck Collaboration.

Comparing the results of the previous two Sections, we showed in detail how models with smaller inflaton decay rates
lead to lower values of $T_{\rm reh}$, $N_*$ and $n_s$, 
whereas the data prefer larger values of $n_s$ and hence $N_*$.
Numerically, in the Starobinsky-like limit models with two-body decays and $y \lesssim 10^{-5}$ as
suggested by the gravitino constraint correspond to $N_* \lesssim 52.7$, whereas the data favour $N_* \gtrsim 50$
at the 68\% CL, with values of $N_* \gtrsim 43$ being allowed at the 95\% CL. The present data therefore tend to
favour models with relatively rapid inflaton decay: $\Gamma_\phi/m \gtrsim 10^{-19}$ corresponding to
$y \gtrsim 2 \times 10^{-9}$ for two-body decays at the 68\% CL, to be compared with the
upper bound $y \lesssim 10^{-5}$ from the gravitino abundance~\footnote{However, the lower limit on
$\Gamma_\phi$ may be relaxed by a factor $\sim 10^6$ (and that on $y$ by a factor $\sim 10^3$)
in non-Starobinsky-like no-scale models.}. It will be interesting
to see how, as the experimental constraints tighten, the experimental noose on $y$ tightens~\footnote{The mild
preference for rapid inflaton decay applies to many models, including the original Starobinsky $R + R^2$
model, no-scale Starobinsky-like models~\cite{ENO6} and Higgs inflation.}

It is interesting to consider specific messages from our analysis for a couple of phenomenological issues,
namely sneutrino inflation and supersymmetry breaking. 

The cosmological upper limit on the gravitino abundance imposes an upper limit on the Yukawa coupling responsible for sneutrino
inflaton decay: $y_\nu \lesssim 10^{-5}$, which is itself an important constraint on realizations of
sneutrino inflation. If the matter field $\Phi$ in the no-scale Wess-Zumino model (\ref{NSWZ}) is
identified with a sneutrino, it must have a trilinear coupling $\lambda \simeq \mu/3$.
Such a trilinear coupling violates $R$-parity, inducing decay of the lightest supersymmetric particle.
However, its lifetime is still much longer than the age of the Universe, and it remains a viable candidate
for cold dark matter. As we have shown here, if $\lambda \gtrsim \mu/3$,
CMB measurements measurements favour $N_* \gtrsim 50$, and hence are on the verge of providing a relevant lower
bound on the coupling $y_\nu$ responsible for sneutrino decay. It will be interesting to see how this squeeze on $y_\nu$ will evolve.

Concerning supersymmetry breaking, we recall that there is a contribution to gaugino masses of the form
\beq
m_{1/2} = \left| \frac{1}{2}e^{G/2}\frac{\bar{f}_{\alpha\beta,T}}{{\rm Re}\,f_{\alpha\beta}} (G^{-1})^T_T G^T \right| = \frac{d_{g,T}}{6}|p-3|m_{3/2}
\eeq
where $p$ is a number of order unity. We saw in Section~3 that strongly-coupled string models with $d_{g,T} = {\cal O}(1)$
would give larger values of the inflaton decay rate $\Gamma_\phi \simeq d_{g, T}^2 m^3/(32\pi)$ (\ref{gammaVV}), 
$T_{\rm reh}, N_*$ and hence $n_s$ than weakly-coupled models with $d_{g,T} = {\cal O}(1/20)$,
as seen in (\ref{weakstrongNstar}). The difference in $N_*$ and hence $n_s$ is not yet significant, but
this connection will also be interesting to watch in the future.

These two examples serve as illustrations how inflationary observables may in the future join the phenomenological
mainstream. These two examples are in the context of specific no-scale supergravity models of inflation, but the
connection has the potential to be more general.

\section*{Acknowledgements}

The work of J.E. was supported in part by the London Centre for Terauniverse Studies
(LCTS), using funding from the European Research Council via the Advanced Investigator
Grant 267352 and from the UK STFC via the research grant ST/L000326/1.
The work of D.V.N. was supported in part by the DOE grant DE-FG03-
95-ER-40917 and in part by the Alexander~S.~Onassis Public Benefit Foundation.
The work of M.A.G.G. and
K.A.O. was supported in part by DOE grant DE-SC0011842  at the University of
Minnesota.

\end{document}